\documentclass[]{aa}

\usepackage{graphicx}
\usepackage{subcaption}
\usepackage{txfonts}
\usepackage{natbib}
\usepackage[colorlinks=true, linkcolor=blue, citecolor=blue, filecolor=blue, urlcolor=blue]{hyperref}
\usepackage{placeins}

\newcommand{\sag}{\textsc{sag}}
\newcommand{\smdpl}{\textsc{smdpl}}

\usepackage[switch]{lineno}
\usepackage{tabularx}
\usepackage{array}

\titlerunning{Cosmic web parameterization}
\authorrunning{Yaryura et al.}

\begin{document} 

   \title{A continuous parameterization of the cosmic web}

   \author{C.Yamila~Yaryura\inst{1,2}\fnmsep\thanks{yamila.yaryura@unc.edu.ar}
          Mario G.~Abadi\inst{1,2}, 
          Noam I.~Libeskind\inst{3},
          Stefan~Gottl\"ober\inst{3},
          Sof\'ia A.~Cora\inst{4,5}
          \and
          Gustavo~Yepes\inst{6,7}
          }

    \institute{CONICET-Universidad Nacional de C\'{o}rdoba, Instituto de
              Astronom\'{i}a Te\'{o}rica y Experimental (IATE), Laprida 854, 
              X5000BGR, C\'{o}rdoba, Argentina
         \and
              Observatorio Astron\'{o}mico, Universidad Nacional de C\'{o}rdoba, Laprida 854, X5000BGR, C\'{o}rdoba, Argentina
         \and
              Leibniz-Institut f\"ur Astrophysik Potsdam (AIP), An der Sternwarte 16, D - 14482, Potsdam, Germany
         \and
              Instituto de Astrof\'isica de La Plata (CCT La Plata, CONICET,UNLP), 
              Observatorio Astron\'omico, Paseo del Bosque,\\ 1900FWA, La Plata, 
              Argentina
         \and
              Facultad de Ciencias Astron\'omicas y Geof\'{\i}sicas, Universidad 
              Nacional de La Plata (UNLP), Observatorio Astron\'omico,\\ Paseo del 
              Bosque, B1900FWA La Plata, Argentina
         \and
              Departamento de F\'{\i}sica Te\'orica M-8, Universidad Aut\'onoma de
              Madrid, Cantoblanco, E-28049 Madrid, Spain    
         \and
              Centro de Investigaci\'on Avanzada en F\'{\i}sica  Fundamental 
              (CIAFF), Universidad Aut\'onoma de Madrid, E-28049 Madrid, Spain
             }

   \date{}

  \abstract
   {The intrinsic properties of galaxies are influenced by their environments, underscoring the environment's critical role in galaxy formation and evolution. Traditionally, these environments are categorized into four fixed classifications: knots, filaments, walls, and voids, which collectively describe the complex organization of galaxies within large-scale structures.
   We propose an alternative description that complements the traditional quadripartite categorization by introducing a continuous framework, allowing for a more nuanced examination of the relationship between the intrinsic properties of galaxies 
   and their environments. 
   This complementary description is applied using one of the most prevalent methodologies: categorization using the eigenvalues of the Hessian matrix extracted from the matter density field. 
   We integrated our findings into a semi-analytical model of galaxy formation, combined with cosmological numerical simulations, to analyze how the intrinsic properties of galaxies are influenced by environmental changes. 
   In our study, we find a continuous distribution of eigenvalue ratios, revealing a clear dependence of galaxy properties on their surrounding environments. This method allowed us to identify critical values at which transitions in the behavior of key astrophysical galaxy properties become evident.}
   \keywords{Cosmology: large-scale structure of Universe -- 
   Galaxies: kinematics and dynamics}
   \maketitle

\section{Introduction}

Large galaxy surveys developed in recent years, such as the 
two degree Field Galaxy Redshift Survey 
(2dFGRS, \citealt{Colless:2003}) and the Sloan Digital Sky Survey 
(SDSS; \citealt{Tegmark:2004}), as well as large numerical 
simulations, show evidence that the distribution of matter in the 
Universe is a nonrandom distribution. The distribution of matter 
on megaparsec scales, traced by galaxies, intergalactic gas 
and dark matter (DM),  
defines an intricate multiscale interconnected 
network known as the cosmic web (\citealt{deLapparent:1986}, 
\citealt{Bond:1996}). 
The cosmic web, which characterizes the large-scale structure of the
Universe, forms through gravitational instability arising from the
initial perturbations in an otherwise homogeneous density field. 
Two prominent features of this web are immediately apparent. 
First, it consists of both high-density and low-density regions. 
Second, the structures within this network are notably nonspherical, 
indicating that gravitational collapse is far from isotropic. 
Thus, the cosmic web emerges from a highly anisotropic gravitational
collapse, producing structures that deviate significantly from
spherical symmetry. 
In other words, the morphology of these structures reflects the
nature of gravitational collapse, which remains ongoing, since many of
these formations have not yet reached virial equilibrium. 
The low-density regions exist as a consequence of the emergence
of high-density structures, and vice versa. 
This interdependence implies that the collapse of one structure 
influences the evolution of others. 
Hence, the structures within the cosmic web are not isolated, 
but rather interconnected, dynamically shaping each other.
\\

While the cosmic web exhibits a continuous and 
interconnected structure, describing it through distinct 
classifications offers a commonly used framework for analyzing 
the diverse environments in which galaxies reside.
The cosmic web is typically categorized into four components: 
knots (or clusters), filaments, walls (or sheets), and voids. 
This classification is generally based on the rate of compression or 
expansion of cosmic material along three orthogonal axes 
 \citep[][among others]{Icke:1991, Cautun:2014, Wang:2017}. 
Various methods have been developed to perform this classification, 
each relying on different physical fields (such as density or velocity) 
and employing different techniques. 
Over the past few decades, these methods have proliferated and can broadly 
be grouped into five categories \citep{Palomino:2024}: 
(i) Hessian-based methods 
\citep[geometric and multi-scale; e.g.,][]{Forero-Romero:2009, Hoffman:2012};
(ii) graph-based methods \citep[e.g.,][]{Alpaslan:2014}; 
(iii) stochastic methods \citep[e.g.,][]{Tempel:2015}; 
(iv) topological methods \citep[e.g.,][]{Sousbie:2011, AragonCalvo:2010}; 
and (v) phase space methods \citep[e.g.,][]{Falck:2012}. 
These approaches rely on the definition of specific 
threshold values to categorize different environments. 
As a result, the classifications obtained naturally reflect the 
methodological design and parameter choices of each technique. 
While this does not undermine their usefulness, it highlights the 
importance of carefully considering how such definitions may influence 
the interpretation of galaxy–environment relations.
\\

The theoretical foundation for applying tidal and deformation 
tensors in the study of the cosmic web was laid by early works such as 
\cite{Shandarin:1989}, which demonstrated how anisotropic structures 
naturally arise from gravitational instability. Subsequent studies 
(e.g., \citealt{Shandarin:2004b}; \citealt{Shandarin:2012})
further emphasized the role of multi-stream flows and the geometry of 
the deformation tensor in shaping filaments and sheets, thereby providing 
the conceptual basis for many of the classification schemes currently in use. 
These developments strongly motivated the adoption of Hessian-based approaches \citep{Olex:2025}, 
where the eigenvalues of the Hessian matrix capture the local dynamical 
behavior of matter and serve as key indicators for identifying different environments within the cosmic web.
\\

The commonly used Hessian-based methods are based on the Hessian 
matrix generated  from a continuous field.
This matrix is diagonalized to estimate its eigenvalues and eigenvectors. 
In this method, a threshold is commonly applied to 
the eigenvalues of the Hessian matrix, typically set to zero 
($\lambda_{\rm th} = 0$).
When the field used is the density field, the sign of a given eigenvalue
at a given point determines whether the gravitational force in the direction
of the corresponding eigenvector is contracting 
(a positive eigenvalue) or expanding (negative).
So, the classification of the cosmic web into four distinct categories
is performed according to the number of eigenvalues exceeding the
established threshold. 
Numerous studies have analyzed the most appropriate value for 
this threshold according to the problem that is being addressed 
address \citep{Hahn:2007a, Forero-Romero:2009, Libeskind:2012, 
Libeskind:2013a, Libeskind:2014b}.
Consequently, alongside the intrinsic characteristics 
of the method, the resulting classification is naturally shaped by 
the specific threshold value adopted for the analysis.
\\

One of the primary aims in developing methods for classifying 
the cosmic web is to determine whether, and how, the cosmic web 
influences the properties of galaxies and their evolution within it. 
There is clear evidence that the properties of DM halos 
and galaxies (such as shape, spin, color, star formation rate, 
and the spatial distribution of satellites) are correlated 
with their large-scale environments 
\citep[][among others]{AragonCalvo:2007, Hahn:2007a, Hahn:2007b, 
Zhang:2009, Hahn:2010, Codis:2012, AragonCalvo:2014, Codis:2015, 
Zhang:2015}. 
To understand how the large-scale structure and the immediate 
environment of a halo or galaxy affect its formation and evolution, 
it may be very useful to use a continuous description of the 
cosmic web.
The traditional discrete classification has undoubtedly 
proven to be a valuable tool, with successful applications across different 
fields, from botany to cosmology, as it provides a practical framework for 
initiating scientific discussions and formulating qualitative insights.
A continuous parameterization does not replace this approach, but rather 
complements it by enabling a more refined and detailed study of the 
dependence of galaxy properties on the surrounding environment.
\\

In this paper we present a simpler and continuous 
description of the cosmic web.
We computed the Hessian matrix from the density field, 
which was derived from the matter distribution traced by the 
positions of DM particles in a cosmological simulation.
We propose describing the cosmic web continuously, 
without categorizing it into separate classes, 
through direct comparisons of the rates of 
compression (or expansion).
This information was derived from the analysis 
of a DM-only simulation, and 
we established the connection to galaxy properties 
by integrating the simulation with a semi-analytic 
model of galaxy formation.
This paper is organized as follows. 
We describe the DM-only simulation and the semi-analytic 
model used in Section \ref{S_methods}.
In Section~\ref{TT}, we describe the method applied to 
classify the cosmic web based on eigenvalues estimated from 
the diagonalized Hessian matrix generated from the spatial 
distribution of the DM particles.
In Section~\ref{Prop}, we analyze the dependence of 
both the galaxy stellar mass and its DM halo mass
on the surrounding environment. 
In Section~\ref{GP}, we examine the influence of 
a galaxy's environment on its main dynamical properties, 
utilizing the cosmic web description presented in this work. 
Finally, in Section~\ref{Conclusions}, we highlight the key 
findings of our study.
\\

\section{Hybrid model of galaxy formation} \label{S_methods}

The galaxy sample is extracted from a catalog generated 
using a hybrid model that combines a semi-analytical approach to 
galaxy formation and evolution with a DM-only cosmological simulation.
Below, we briefly describe the main aspects of this model.
\\

\subsection{Dark matter cosmological simulation}
\label{sec:simSMDPL}

We used the \smdpl~DM-only cosmological
simulation\footnote{https://www.cosmosim.org/metadata/smdpl/}, which follows the 
evolution of $3840^3$ particles from redshift $z = 120$ to $z = 0$, 
within a (relatively) small volume (a periodic box 
of side-length of $400\,{\rm Mpc}\,h^{-1}$).
This large number of particles within such a volume 
reaches a mass resolution 
of $9.63\times10^7 \,{\rm M}_{\odot}\,h^{-1}$ per DM 
particle (see \citealt{Klypin:2016} for more details).
\smdpl~cosmological parameters are given by a 
flat $\Lambda$ cold DM model 
consistent with Planck measurements: 
\mbox{$\Omega_{\rm m}$ = 0.307}, 
\mbox{$\Omega_{\rm B}$ = 0.048}, \mbox{$\Omega_{\Lambda}$ = 0.693}, 
\mbox{$\sigma_{8}$ = 0.829},
\mbox{$n_{\rm s}$ = 0.96}, and \mbox{$h$ = 0.678} \citep{Planck:2014}.
\\

The \textsc{Rockstar} halo finder \citep{Behroozi_rockstar} is used 
to identify DM halos keeping just overdensities with at 
least $N_\text{min}$~=~20~DM particles. 
There are two classifications of DM halos: main host halos
(detected over the background density) and subhalos 
(that lie inside other DM halos). 
From these halos, \textsc{ConsistentTrees} \citep{Behroozi_ctrees} 
constructs merger trees, by linking halos and subhalos 
forward and backward in time to progenitors and 
descendants, respectively. 
\\

\subsection{Semi-analytic model of galaxy formation, \sag~}

We used the latest version of the semi-analytic model 
\sag~described in \cite{Cora:2018}, which is
based on the model presented by \cite{Springel:2001}. 
This is an updated and improved version, including the main physical 
processes required for galaxy formation; we refer the reader to 
\cite{Cora:2018} for a detailed 
and exhaustive description of the model.
Each \sag~galaxy populates a DM halo of the simulation in 
such a way that central galaxies correspond to
main host halos while satellite galaxies are hosted by subhalos, 
according to the information provided by the merger trees.
These satellites are defined as orphans when the mass of their 
DM substructures is no longer detected by the halo finder.
However, in this work we focused only on central galaxies.
To regulate the physical processes involved in the \sag~model, a set of 
free parameters are employed. 
These parameters are calibrated using a set of observed galaxy
properties.
The particle swarm optimization technique \citep{Kennedy:1995} and the 
observational data presented in \cite{Knebe:2018} were used to 
select the best-fitting values for the free parameters of \sag~for 
the \smdpl~simulation \citep{Ruiz:2015}. 
The reader can refer to Table 1 in \cite{Yaryura:2023}, where these 
best-fitting values are listed.
\\

\section{Tidal field estimation} \label{TT}

In recent decades, numerous methods have been 
developed to classify the cosmic web \cite[see][for a review]{Libeskind2018}. 
Many of these methods determine the Hessian matrix from the density, 
velocity, or potential field using a fixed finite grid. 
This matrix is diagonalized to determine its eigenvectors 
and eigenvalues, which provide information about the principal 
directions of local collapse or expansion.
In practice, when estimating the Hessian matrix from the density, 
what is done is constructing a discrete density field from the discrete 
distribution of particles using the cloud-in-cell 
technique with a given number of grids, $N^3_{\rm grid}$. 
A spherically symmetric Gaussian window function is applied 
to smooth this density field, obtaining a smoothed, 
continuous density field, $\rho(x)$. 
The gravitational potential originated by this distribution obeys the 
(rescaled) Poisson equation:

\begin{equation}
    \nabla^{2} \phi = \delta,
\end{equation}

\noindent where $\phi$ is the gravitational potential rescaled by $4 \pi G \bar{\rho}$, $\bar{\rho}$ is 
the mean cosmological density, and the density contrast ($\delta$) is given by $\delta = \rho / \bar{\rho} -1$.
\\

Furthermore, the Hessian of the gravitational potential $\phi$ is sometimes 
called the deformation tensor (see for example \citealt{Forero-Romero:2009}) 
and is given by

\begin{equation}
    T_{\alpha, \beta} = \frac{\partial^2 \phi}{\partial r_{\alpha} \partial r_{\beta}},
\end{equation}

where $\alpha$ and $\beta$ represent the spatial coordinates x, y and z. 
By diagonalizing this matrix, its eigenvalues $a>b>c$ are estimated, and 
they satisfy the following relationship:

\begin{equation}
    \delta = a + b + c.
\end{equation}

These eigenvalues represent the directional changes in the 
gravitational field's intensity, offering insights into regions of 
compression or expansion. 
Specifically, the sign of each eigenvalue at a given point reveals the
nature of the gravitational influence along the direction of its
associated eigenvector: a positive value indicates contraction, 
while a negative value indicates expansion.
Starting from the case where all three eigenvalues are positive, 
to the case where all three are negative, and passing through the
intermediate cases, the density decreases.
\\

From these eigenvalues, the cosmic web is commonly classified into four 
distinct categories: knots, filaments, walls and voids, according to the 
number of eigenvalues greater than a certain threshold 
value ($\lambda_{\rm th}$):

\begin{itemize}
  \item {Knot: if $\lambda_{\rm th} < c$ }
  \item {Filament: if $ c < \lambda_{\rm th} < b $ }
  \item {Wall: if $ b < \lambda_{\rm th} < a $ }
  \item {Void: if $ a < \lambda_{\rm th}$}
\end{itemize}

So, this usual cosmic web classification mainly depends on two parameters: 
the smoothing scale and the threshold value $\lambda_{\rm th}$ selected.
Although $\lambda_{\rm th} = 0$ is the most commonly used value 
\citep{Hahn:2007a}, there are numerous studies about the most suitable 
value for this threshold value \citep{Forero-Romero:2009, Libeskind:2012, 
Libeskind:2013a, Libeskind:2014b}.
Several of these works recommend different values depending on the 
specific research focus \citep{Forero-Romero:2009}. 
For example, in our study of associations of dwarf galaxies within 
the same simulation, 
we applied this classification using $\lambda_{\rm th} = 0$ \citep{Yaryura:2023}. 
Consequently, this classification is highly sensitive to the selected 
threshold value.
\\

The complex and continuous nature of the large-scale 
structure of the Universe motivates the development of alternative 
methods for characterizing the cosmic web. 
In this context, we propose a simple and continuous description 
based on directly examining the ratios of the eigenvalues of the 
Hessian matrix derived from the matter distribution.
We estimated the eigenvalues from the tidal 
field \citep[e.g.,][]{Hahn:2007a} 
of the DM particle distribution of the parent simulation \smdpl~using 
the method presented in \cite{Libeskind:2013a}.
Based on the position of the DM particles in the 
\smdpl~simulation, we used a grid of $400^3$ cells 
and three different specific smoothing lengths 
$l=1,\,2$ or 4 $\,{\rm Mpc}\,h^{-1}$ 
to estimate the three eigenvalues for each cell. 
So, to each cell corresponds: $a>b>c$. 
\\

The importance of describing the cosmic web lies, among other factors, 
in studying where galaxies are preferentially located and how 
their main properties are influenced by the surrounding environment.
The sample of galaxies studied in this work is obtained from
a set of semi-analytical galaxies
built by enforcing a minimum stellar and halo mass of  
$M_{\star} = 10^{9}\,{\rm M}_{\odot}\,h^{-1}$, and 
$M_{200} = 10^{11} \, {\rm M}_{\odot}\,h^{-1}$ 
(equivalent to $\sim 1000$ DM particles), respectively.
The final sample has a total of $1\,688\,127$ 
well resolved, semi-analytical central galaxies, with stellar 
masses in the range 
$9.0 < {\rm log_{10}}(M_{\star}[{\rm M}_{\odot}\,h^{-1}]) < 12.9$, 
hosted by DM halos with masses ranging between 
hosted by DM halos with masses ranging between 
$11.0 < {\rm log_{10}}(M_{200}[{\rm M}_{\odot}\,h^{-1}]) < 15.2$.
The existence of central galaxies with stellar masses exceeding 
$10^{12}\,M_\odot\,h^{-1}$ stems from the current version of \sag's inability 
to accurately fit the massive end of the stellar mass function at $z = 0$. 
This leads to an excess of galaxies with stellar mass 
\mbox{$M_{\star} \gtrsim 1.3 \times 10^{11} \, M_\odot \, h^{-1}$}, 
as already indicated in \cite{Cora:2018}, where possible causes are proposed.
\\

\begin{figure}
    \includegraphics[width=.47\textwidth]{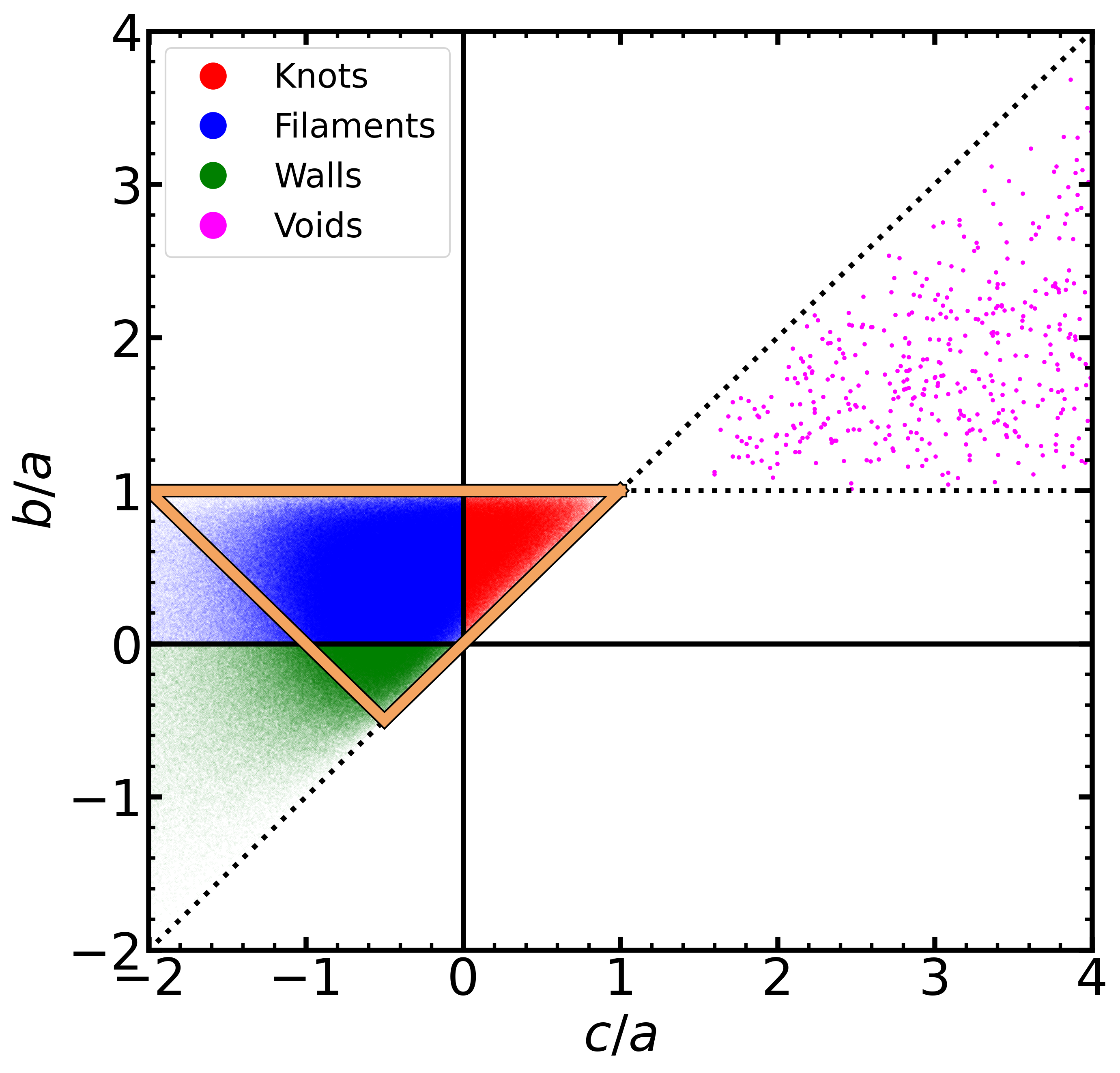}
    \caption{Relationship between the eigenvalue ratios $b/a$ and $c/a$ for central galaxies, according 
    to conventional classification of the cosmic web, using the threshold 
    $\lambda_{\rm th} = 0$. 
    Each point corresponds to one galaxy, which is color-coded according to 
    the eigenvalues associated with the cell it occupies as indicated by the 
    legend.
    The size of the magenta points has been arbitrarily increased for visibility, as they are very sparse.
    The dotted black lines delineate the regions that the eigenvalue ratios may occupy. 
    The horizontal dotted black line at \mbox{$b/a=1$} indicates that \mbox{$b/a<1$} 
    whenever \mbox{$a>0$} and \mbox{$b/a>1$} only when \mbox{$a<0$}.
    The positively sloped dotted black line indicates the condition \mbox{$a>b>c$}, 
    which implies \mbox{$b/a>c/a$} if \mbox{$a>0$} and \mbox{$b/a<c/a$} if $a<0$.
    The area inside (outside) the orange triangle corresponds to $\delta > 0$ ($\delta < 0$).}
    \label{fig:kfwv_color}
\end{figure}

\begin{figure*}
    \centering
    \begin{subfigure}[b]{0.5\textwidth}
        \centering
        \includegraphics[width=\textwidth]{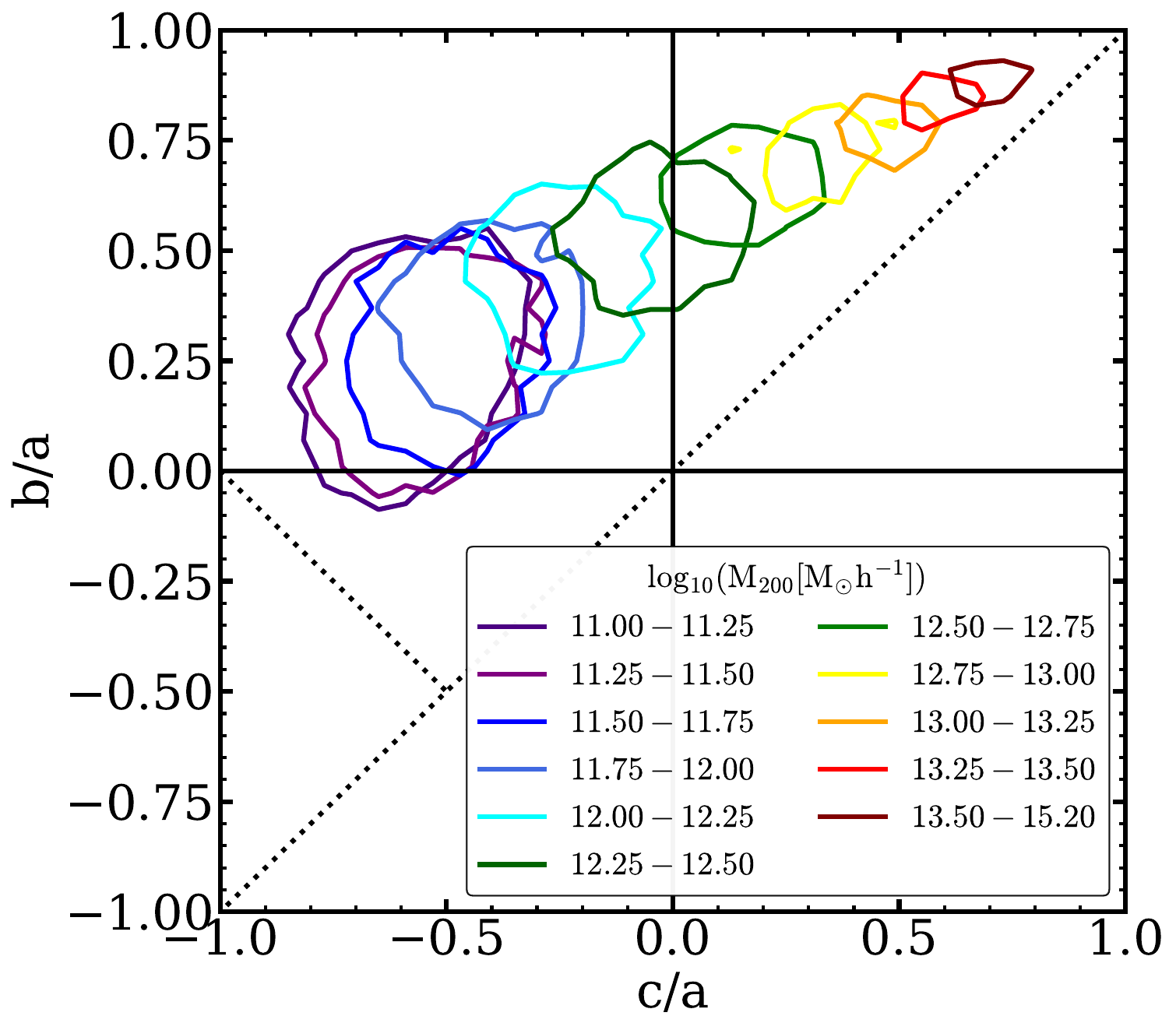}
        \label{fig:mass200_a}
    \end{subfigure}%
    \begin{subfigure}[b]{0.5\textwidth}
        \centering
        \includegraphics[width=\textwidth]{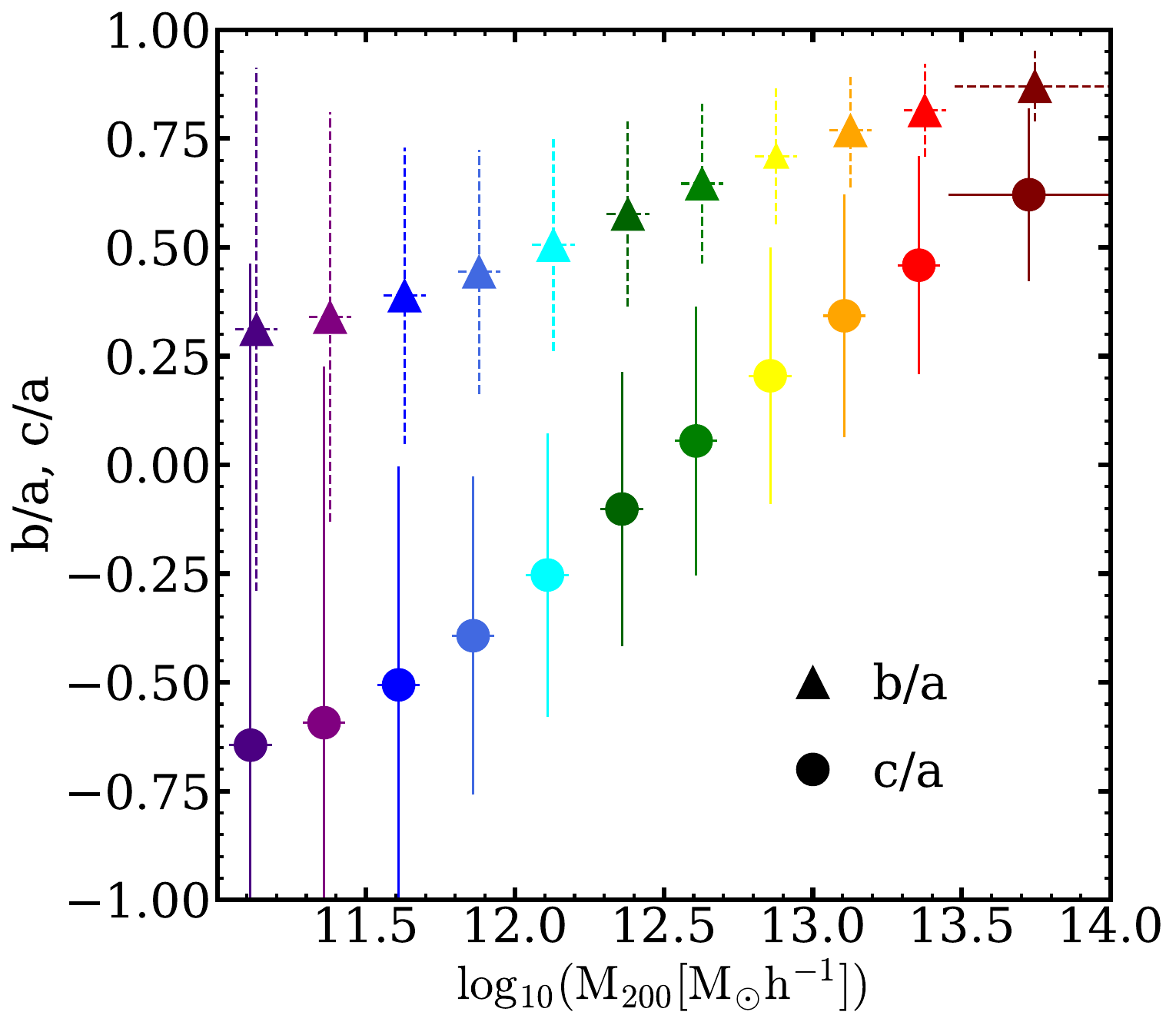}
        \label{fig:mass200_b}
    \end{subfigure}
    \caption{
    Eigenvalue ratios, $b/a$ and $c/a$, as a function of the logarithm of the 
    halo mass ($M_{200}$). 
  Left panel: $b/a$ as a function of $c/a$ in bins of the logarithm of halo mass. 
  This panel shows a zoomed-in view of the region displayed in Fig. 
  \ref{fig:kfwv_color}, centered at origin. 
  The dotted black lines delineate the regions that the eigenvalue ratios may occupy. 
  The upper limit of the graph, \mbox{$b/a=1$}, indicates that \mbox{$b/a<1$} 
  whenever \mbox{$a>0$}.
  The positively sloped dotted black line indicates the condition \mbox{$a>b>c$}, 
  which implies \mbox{$b/a>c/a$} if \mbox{$a>0$} and \mbox{$b/a<c/a$} if $a<0$.
  Right panel: Median values of $b/a$ (triangles) and $c/a$ (circles) for equal bins 
  of the logarithm of the halo mass, as a function of the logarithm of the halo mass. 
  In both panels, each color corresponds to each halo mass bin, as indicated by the legend.}
    \label{fig:mass200}
\end{figure*}

So, the environment of each simulated galaxy extracted from the \sag~model 
is parameterized by the eigenvalues of the cell in which it is located. 
To estimate these eigenvalues, we applied a spherically symmetric 
Gaussian window function to smooth the density field, resulting in a 
continuous smoothed density field $\rho(x)$. 
This procedure was performed using different smoothing lengths.
We verified that the results obtained in this work do not change
significantly when we use different smoothing lengths
($l=1,\,2,$ or 4 $\,{\rm Mpc}\,h^{-1}$). 
Therefore, for simplicity and clarity, the results of this paper
will be presented only for the smoothing length $l=1 \,{\rm Mpc}\,h^{-1}$, 
which is a sufficient volume to analyze the environment of galaxies.
In the Appendix \ref{App}, we present the differences arising 
from the use of three distinct smoothing scales ($l=1,\,2,$ or 4 $\,{\rm Mpc}\,h^{-1}$), 
as well as the differences in estimating eigenvalues derived 
from either the DM density field or the shear velocity field.
\\

Considering the previously described sample of galaxies, 
we analyzed the properties of the eigenvalues by 
examining the ratios between the smallest and the largest eigenvalue 
($c/a$), and the intermediate and the largest eigenvalue ($b/a$), 
in Fig.~\ref{fig:kfwv_color}. 
Each point corresponds to one galaxy, whose eigenvalues are 
associated with the cell it occupies.
Unlike, for example, shape measurements (where all eigenvalues are positive 
and the ratios are naturally bounded), here the eigenvalues can take both 
positive and negative values, resulting in unbounded ratios. 
As a consequence, the axes in this figure are truncated at arbitrary limits, 
beyond which the density of points becomes negligible. 
This plot visually demonstrates that the distribution of eigenvalue 
ratios is continuous, with no obvious natural divisions or discrete 
categories emerging in this plane. 
For reference, we have color-coded the data according to the standard 
discrete classification of the cosmic web into knot (red points), 
filament (blue points), wall (green points), and void (magenta points). 
For this classification, we used the threshold 
$\lambda_{\rm th} = 0$.
The size of the magenta points, corresponding to galaxies located in 
regions that are expanding along all three orthonormal axes (voids), 
has been arbitrarily increased for visibility, as they are very sparse.
In this figure, it is important to distinguish different regions 
according to the values of the eigenvalues, which imply regions of 
compression or expansion. 
When the three eigenvalues are positive ($0<b/a<1$ and $0<c/a<1$), 
implies regions that are collapsing on the three axes.
When two of them are positive and the smallest eigenvalue is negative 
($0<b/a<1$ and $c/a<0$), corresponds to regions that 
are collapsing on two axes but expanding in the third one.
When only the largest eigenvalue is positive and the other two are negative 
($b/a<0$ and $c/a<0$), suggests regions that 
are expanding on two axes but collapsing in the third one.
Finally, when all three eigenvalues are negative ($b/a>1$ and $c/a>1$), 
it notices regions that are expanding in the three directions, 
indicating the lowest dense regions of the Universe. 
The dotted black lines delineate the regions that the eigenvalue ratios may 
occupy.
The horizontal dotted black line at \mbox{$b/a=1$} indicates that \mbox{$b/a<1$} 
whenever \mbox{$a>0$}. 
The only case where \mbox{$b/a>1$} is when \mbox{$a<0$}, which implies that 
the other two eigenvalues, b and c, are also negative, corresponding to the region 
known as voids.
The positively sloped dotted black line indicates the condition 
\mbox{$a>b>c$}, which implies \mbox{$b/a>c/a$} if \mbox{$a>0$} (the case for knots, 
filaments, and walls) and \mbox{$b/a<c/a$} if $a<0$ (voids).
Lastly, the area inside the triangle demarcated by orange lines corresponds to 
$\delta > 0$, while the region outside this triangle corresponds to $\delta < 0$.
In this figure, it is evident that the separation between knots, filaments, 
and walls depends exclusively on the chosen value of $\lambda_{\rm th}$; 
in this case, \mbox{$\lambda_{\rm th} = 0$}.
\\

\begin{figure*}
    \centering
    \begin{subfigure}[b]{0.5\textwidth}
        \centering
        \includegraphics[width=\textwidth]{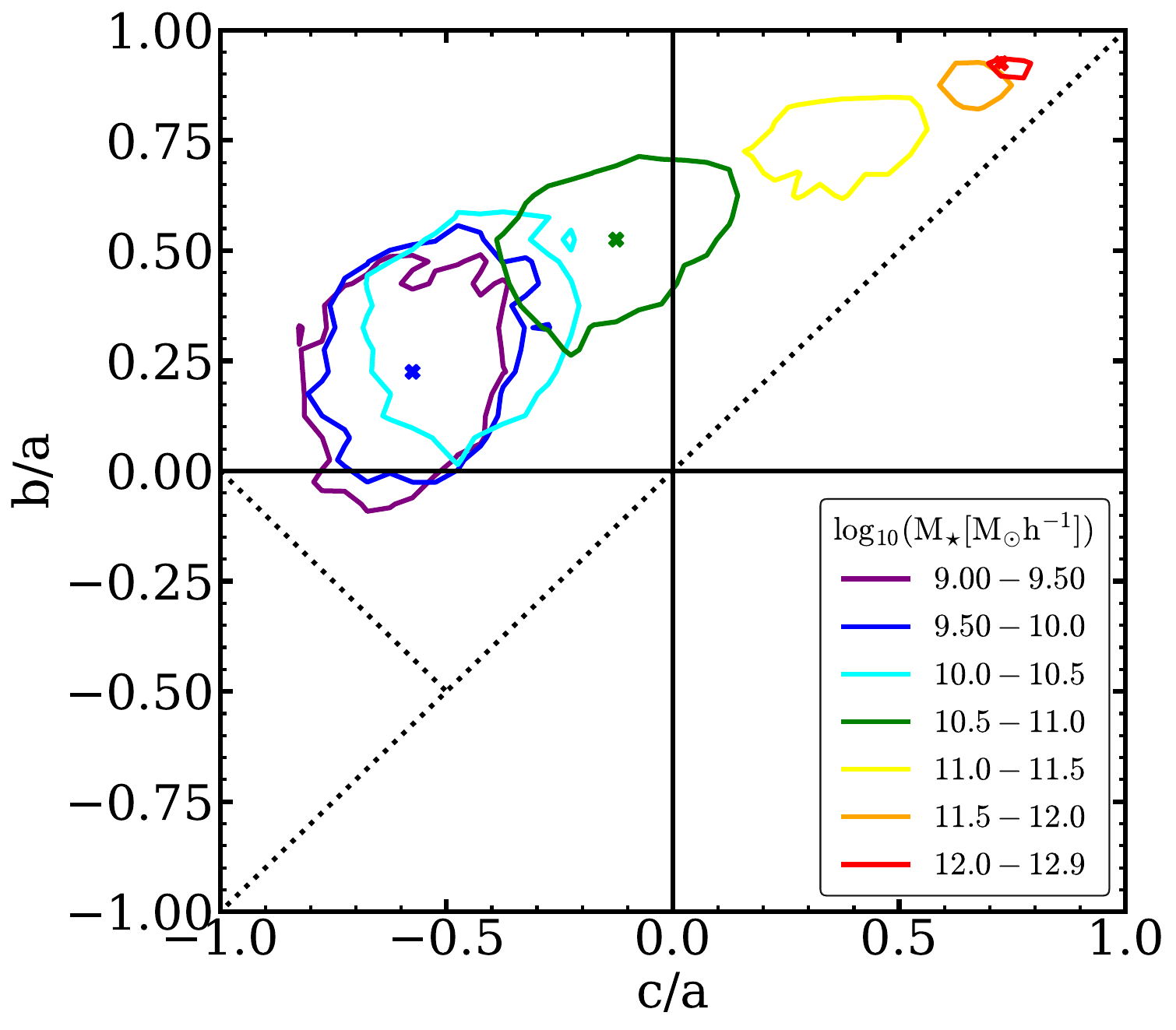}
        \label{fig:smass_a}
    \end{subfigure}
    \begin{subfigure}[b]{0.48\textwidth}
        \centering
        \includegraphics[width=\textwidth]{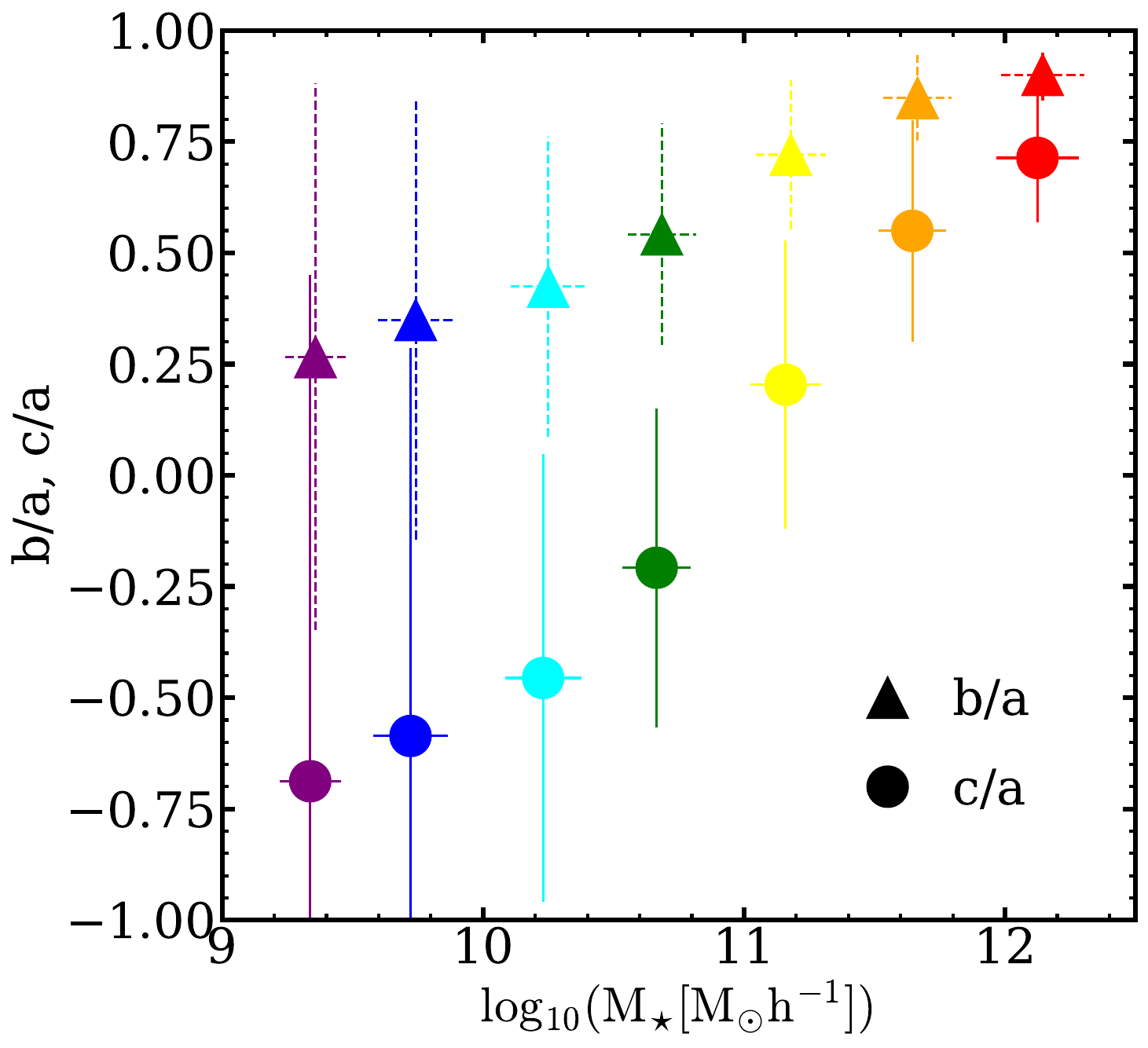}
        \label{fig:smass_b}
    \end{subfigure}
  \caption{Eigenvalues ratios, $b/a$ and $c/a$, as a 
  function of the logarithm of the stellar mass ($M_{\star}$). 
  Left panel: Eigenvalues ratios, $b/a$ vs. $c/a$, in bins of the logarithm of 
  stellar mass. 
  Right panel: Median values of $b/a$ (triangles) and $c/a$ (circles) for 
  equal bins of the logarithm of the stellar mass, as a 
  function of the logarithm of the stellar mass. 
  In both panels, each color corresponds to each stellar mass bin, as indicated by the legend.}
    \label{fig:smass}
\end{figure*}

\section{Dependence of galaxy properties on the environment} \label{Prop}

Several studies have been conducted to comprehend the influence of 
the environment on galaxy formation and evolution, proving that 
the environment plays a fundamental role in both the formation 
and evolution of galaxies.
There is ample evidence that the properties 
of galaxies and galaxy systems depend on their environment 
\citep[][among others]{Dressler:1980,Kauffmann:2004,Blanton:2005, 
OMill:2008,Peng:2010,Wetzel2012,Zheng:2017,Wang2018,Duplancic:2020}. 
For example, it is well known that stellar mass is strongly correlated 
with the environment. 
The stellar mass distribution of galaxies shifts by almost a factor of 
2 toward higher masses between low- and high-density 
regions \citep{Kauffmann:2004}. 
Analyzing the morphology, elliptical galaxies are more frequently 
located in denser regions, while spiral galaxies are more common 
in the field \citep{Dressler:1980, Kuehn:2005, Tempel:2011, Pfeffer:2023}. 
Similar trends are observed for colors, 
star formation rate (SFR), and the ages 
of galaxies \citep{Blanton:2005}; in denser environments, there is a higher 
proportion of red and passive galaxies for a given stellar mass 
\citep{Baldry:2006, Wetzel2012, Wang2018}. 
\cite{Kauffmann:2004} conclude that the galaxy property most sensitive to 
the environment is the specific star formation rate (sSFR; defined as SFR/$M_{\star}$). 
They show that for galaxies with stellar masses in the 
range $10^{10} - 3 \times 10^{10} M_{\odot}$, 
the median sSFR decreases by more than a factor of 10 as the 
population shifts from predominantly star-forming at low densities 
to predominantly passive at high densities. 
This decrease is less marked but still significant for high-mass galaxies.
\cite{Bamford:2009} analyze the relationships between galaxy morphology, color, 
environment and stellar mass. 
They conclude that color and morphology are both sensitive to stellar mass.
However, at fixed stellar mass, while color is also highly sensitive to 
the environment, morphology displays much weaker environmental trends. 
In any case, despite all the studies carried out so far, understanding the effects 
of the environment on galaxy properties remains a complex and active area of research. 
\\

In light of the numerous studies highlighting the role of the 
environment in galaxy formation and evolution, we aim to revisit these studies 
using a complementary approach—one that relies on a continuous, rather than discrete, 
characterization of the large-scale environment. 
In this framework, we examined how various galaxy 
properties correlate with their surrounding environment, as described 
by this continuous method.
As a first step, we analyzed the values of the eigenvalue 
ratios $b/a$ and $c/a$ to identify the regions where galaxies 
preferentially reside, 
as a function of their DM halo mass.
The left panel of Fig.~\ref{fig:mass200} shows the 
relation between the ratios of eigenvalues, $b/a$ as a function of $c/a$, 
for the galaxies sample, in bins of the logarithm of their 
DM halo mass ($M_{200}$).
Each contour corresponds to each mass bin (color indicated by the legend), 
centered at the maximum of each sample indicating the $80$ per cent.
It is evident that galaxies hosted by more massive halos are located in 
denser regions, where both $b/a$ and $c/a$ increase, 
approaching values close to unity in regions where $\delta > 0$. 
Conversely, galaxies residing in less massive DM halos are 
found in regions where the values of $b/a$ and $c/a$ decrease, 
in some cases even reaching negative values.
While there is a clear dependence of DM halos mass on 
the environment where galaxies inhabit, this variation is smooth. 
Another way to analyze this dependence is by studying 
the variation of the ratios of eigenvalues as a function of the DM 
halos mass, as shown in the right panel of Fig.~\ref{fig:mass200}. 
This figure presents the median values of $b/a$ (triangles) and 
$c/a$ (circles) in bins of the logarithm of the DM halos mass, 
where each color corresponds to a different DM halos mass bin, 
as indicated by the legend. 
The median value of $b/a$ increases as we consider galaxies 
with more massive DM halos. 
As we consider environments of galaxies hosted by 
less massive DM halos, 
around $M_{200} \sim 10^{11}\,{\rm M}_{\odot}\,h^{-1}$, 
toward environments of galaxies hosted by 
more massive DM halos
($M_{200} \sim 10^{13.5}\,{\rm M}_{\odot}\,h^{-1}$), 
$b/a$ increases from $b/a \sim 0.30$ to $b/a \sim 0.85$. 
Despite this mass dependence, the median 
values of $b/a$ remains positive within the considered mass range. 
On the other hand, the median values of $c/a$ also increase as we 
consider galaxies with more massive DM halos. 
The variation in $c/a$ values with mass is significantly greater 
than the variation in $b/a$. 
Median values of $c/a$ are negative for galaxies within less 
massive DM halos and become positive for those within 
more massive halos. 
The median $c/a$ is approximately $c/a \sim -0.65$ for galaxies hosted 
by less massive halos ($M_{200} \sim 10^{11}\,{\rm M}_{\odot}\,h^{-1}$). 
The transition from negative to positive $c/a$ values occurs at 
$M_{200} \sim 5\times 10^{12}\,{\rm M}_{\odot}\,h^{-1}$, reaching values 
of approximately $c/a \sim 0.60$ for galaxies hosted by more massive 
DM halos ($M_{200} \sim 10^{13.5}\,{\rm M}_{\odot}\,h^{-1}$).
\\

In summary, both the right and left panels of Fig.~\ref{fig:mass200} 
clearly show that the relationship between DM halos mass and the environment 
where galaxies are situated is continuous.
We notice that the mass at which the median value 
of $c/a$ transitions from negative to positive is tantalizing close to the 
characteristic mass discussed by \cite{Dekel:2006}. 
They examine the impact of a critical halo mass in governing 
the bimodality of galaxy properties by modulating gas inflow, 
star formation, and feedback processes, ultimately driving the transition 
from blue star-forming disks to red passive spheroids.
\\

\begin{figure*}
    \centering
    \begin{subfigure}[b]{0.34\textwidth}
        \centering
        \includegraphics[width=\textwidth]{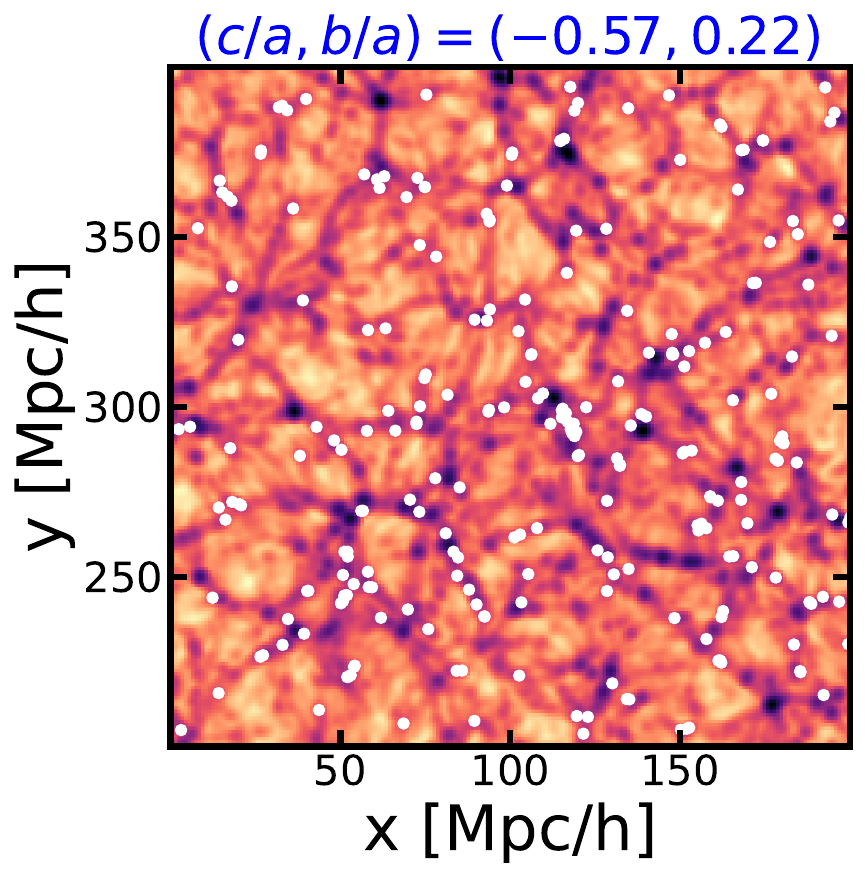}
        \label{fig:sd_low}
    \end{subfigure}%
    \begin{subfigure}[b]{0.28\textwidth}
        \centering
        \includegraphics[width=\textwidth]{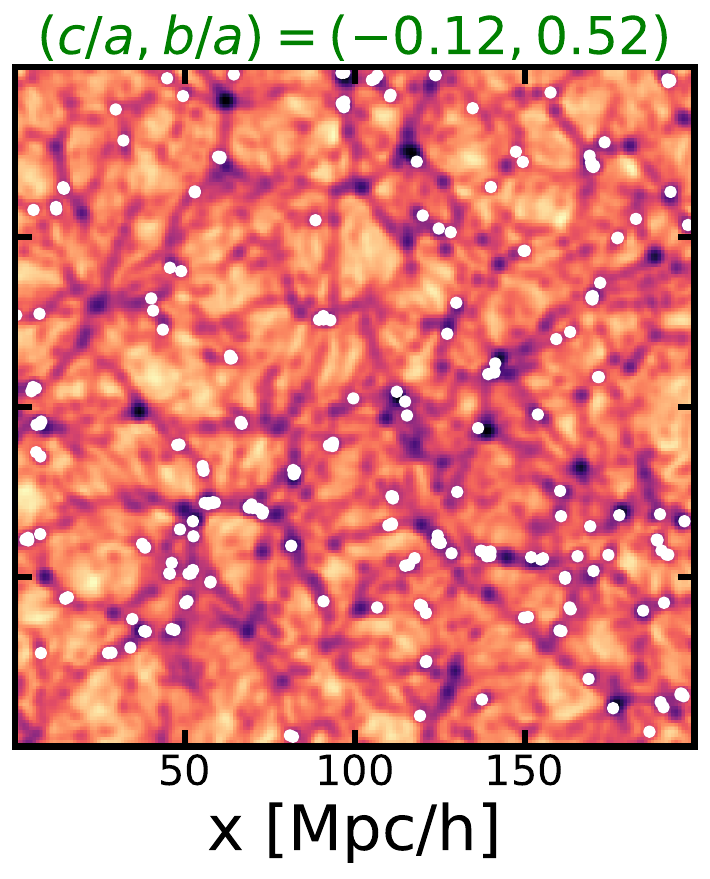}
        \label{fig:sd_med}
    \end{subfigure}
    \begin{subfigure}[b]{0.37\textwidth}
        \centering
        \includegraphics[width=\textwidth]{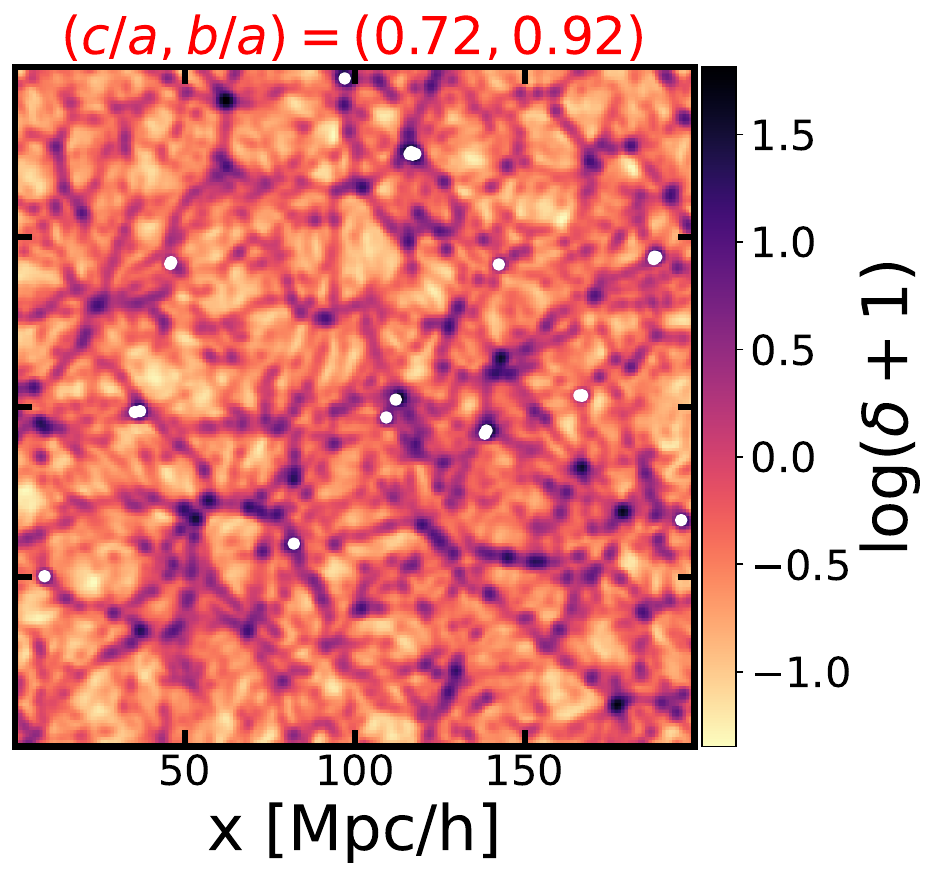}
        \label{fig:sd_high}
    \end{subfigure}    
    \caption{Spatial distribution of central galaxies, in a slice of 
$1 \,{\rm Mpc}\,h^{-1}$ in a box of the \smdpl~simulation. 
The background shows the logarithm of the density contrast estimated 
from DM particles, where the colors correspond to the values indicated 
in the color bar.
White filled circles represent galaxies located in environments where 
the values of their eigenvalue ratios $b/a$ and $c/a$ (corresponding 
to the cell in which they are located) fall within a circle of radius 
$0.1 \,{\rm Mpc}\,h^{-1}$ centered on the peaks of three of the contours 
in the left panel of Fig.~\ref{fig:smass} marked with crosses.
The panels correspond to the blue (left), green (middle), 
and red crosses (right).
Above each panel, the ($c/a$, $b/a$) values are indicated for each cross.}
    \label{fig:spatial_dist}
\end{figure*}

Figure~\ref{fig:smass} shows the relationship between the eigenvalue 
ratios, $b/a$ and $c/a$, similar to Fig.~\ref{fig:mass200}, 
but in this case, considering bins in the logarithm of stellar mass of galaxies 
($M_{\star}$).
As expected, the variation of the environment according to $M_{\star}$ 
is consistent with the previously presented results for the DM 
halos mass ($M_{200}$).
In the right panel of Fig.~\ref{fig:smass}, more massive galaxies are located 
preferably in denser environments where the median values of $b/a$ and $c/a$ 
are both positive.
As we consider less massive galaxies, both the values of $b/a$ and 
$c/a$ decrease, indicating that these galaxies reside in less dense 
environments.
As we consider environments of less massive galaxies, around 
$M_{\star} \sim 10^{9}\,{\rm M}_{\odot}\,h^{-1}$, 
toward environments of more massive galaxies 
($M_{\star} \sim 10^{12}\,{\rm M}_{\odot}\,h^{-1}$), 
$b/a$ increases from $b/a \sim 0.25$ to $b/a \sim 0.90$. 
Despite the variation in $b/a$, its median values always remain positive, 
indicating that the two major axes are always positive. 
In contrast, when analyzing the variation of $c/a$ with respect to stellar mass, 
we observe that environments of less massive galaxies correspond to negative 
values of $c/a$. 
The median $c/a$ is approximately $c/a \sim -0.75$ for the 
least massive galaxies ($M_{\star} \sim 10^{9}\,{\rm M}_{\odot}\,h^{-1}$). 
These values become positive for intermediate stellar mass galaxies, 
around $M_{\star} \sim 10^{11}\,{\rm M}_{\odot}\,h^{-1}$, reaching values of 
approximately $c/a \sim 0.75$ for the most massive galaxies 
($M_{\star} \sim 10^{12}\,{\rm M}_{\odot}\,h^{-1}$). 
As highlighted in Fig.~\ref{fig:mass200}, it is evident 
that the variation in $c/a$ is significantly greater 
than the variation in $b/a$ as a function of stellar mass.
\\

We display the spatial distribution of our galaxies sample
in Fig.~\ref{fig:spatial_dist}, in a slice of 
$1 \,{\rm Mpc}\,h^{-1}$ thickness in a box of the \smdpl~simulation.
The background is the same for all three panels and shows the 
logarithm of the density contrast estimated from DM particles, 
where colors are indicated by the color bar.
The white filled circles represent galaxies located in environments where 
the values of their eigenvalue ratios $b/a$ and $c/a$ (corresponding 
to the cell in which they are located) fall within a circle of radius 
$0.1 \,{\rm Mpc}\,h^{-1}$ centered on the peaks of three of the contours 
in Fig.~\ref{fig:smass}.
Each of these centers is marked with a cross in the left panel of 
Fig.~\ref{fig:smass}. 
Left panel corresponds to blue cross ($c/a=-0.57$, $b/a=0.22$), 
middle panel to green cross ($c/a=-0.12$, $b/a=0.52$) 
and right panel to red cross ($c/a=0.72$, $b/a=0.92$).
From left to right, the galaxies selected are located in increasingly 
dense regions.
Moreover, considering the trend observed in Fig.~\ref{fig:smass} and 
Fig.~\ref{fig:mass200}, as we move toward higher densities, 
the galaxies become more massive, which explains their lower abundance.
Less massive galaxies are significantly more numerous than 
their massive counterparts, so the number of galaxies decreases 
from left to right across the three panels.
\\

While this work presents results based solely on cosmological 
simulations, the proposed description of the cosmic web can also be applied 
to observational data. 
Next-generation galaxy surveys, such as the Dark Energy Spectroscopic Instrument 
(DESI; \citealt{DESI:2016a, DESI:2016b}) and the Vera C. Rubin Observatory’s 
Legacy Survey of Space and Time (LSST; \citealt{Ivezic:2019}), among others, 
will map an unprecedented volume of the Universe. 
These ambitious efforts will enable significant progress in our understanding 
of cosmic structure and evolution, underscoring the need for robust and 
efficient techniques to reconstruct the 3D DM density, 
velocity, and tidal fields from extensive galaxy survey datasets.
A number of studies have addressed the reconstruction of these fields 
from observational data \citep{Fisher:1995, Zaroubi:1995, Schmoldt:1999, 
Mathis:2002, Erdogdu:2004, Wang:2009, Pen:2012, Wang:2013, Wang:2024, 
Lamman:2024}. 
Among the most recent developments, \cite{Shi:2025} introduced a robust 
machine-learning-based framework for reconstructing the DM density, 
velocity, and tidal fields from DESI-like bright galaxy samples. 
Their approach builds upon the deep-learning method for recovering the 
DM density field from the redshift-space distribution of DM 
halos presented in \cite{Wang:2024}, incorporating enhancements that 
account for realistic observational effects, such as geometric selection, 
flux-limited samples, and redshift space distortions.
While the model developed by \cite{Shi:2025} is trained and validated using 
mock samples of the DESI bright galaxy survey, its design allows direct 
applications to real observational data.
To do so, one requires a galaxy redshift survey with 
spectroscopic redshifts, angular positions (RA, Dec), and photometry 
(e.g., z-band magnitudes), along with a well-characterized survey footprint 
and angular mask to account for regions of incompleteness or contamination. 
The input to the model is the 3D galaxy overdensity field in redshift space, 
constructed from the observed galaxy distribution using the assumed cosmology, 
without requiring knowledge of peculiar velocities or DM properties. 
This enables the framework to infer the underlying DM density, velocity, 
and tidal fields directly from observable quantities.
Their results demonstrate that the proposed framework accurately captures 
the large-scale distribution and dynamics of DM while effectively 
addressing key observational systematics. 
These advances provide a reliable and versatile tool for analyzing data 
from current and upcoming galaxy surveys. 
Once the dark tidal fields have been reconstructed from observational 
data using any of the currently available methods, the same procedure 
described earlier can be applied to estimate the eigenvalues of the 
tidal field and provide a continuous description of the cosmic web.
\\

\section{Galaxies properties} \label{GP}

Galaxies properties, including stellar mass, color, 
and SFR, are known to exhibit strong correlations 
with one another, which can be influenced by environmental factors. 
In particular, there is a significant dependence of galaxy properties 
on their stellar mass. 
Less massive galaxies tend to be bluer and have higher SFRs compared 
to more massive galaxies, which are redder and exhibit lower SFRs. 
Moreover, the way in which galaxy properties are influenced by the 
environment is an ongoing area of study. 
As previously mentioned, one of the most commonly used methods 
for studying the variation of galaxy properties 
according to their environment is to classify the environment into 
discrete categories, defined based on different methods. 
\\

\begin{figure*}
    \centering
    \includegraphics[width=0.73\textwidth]{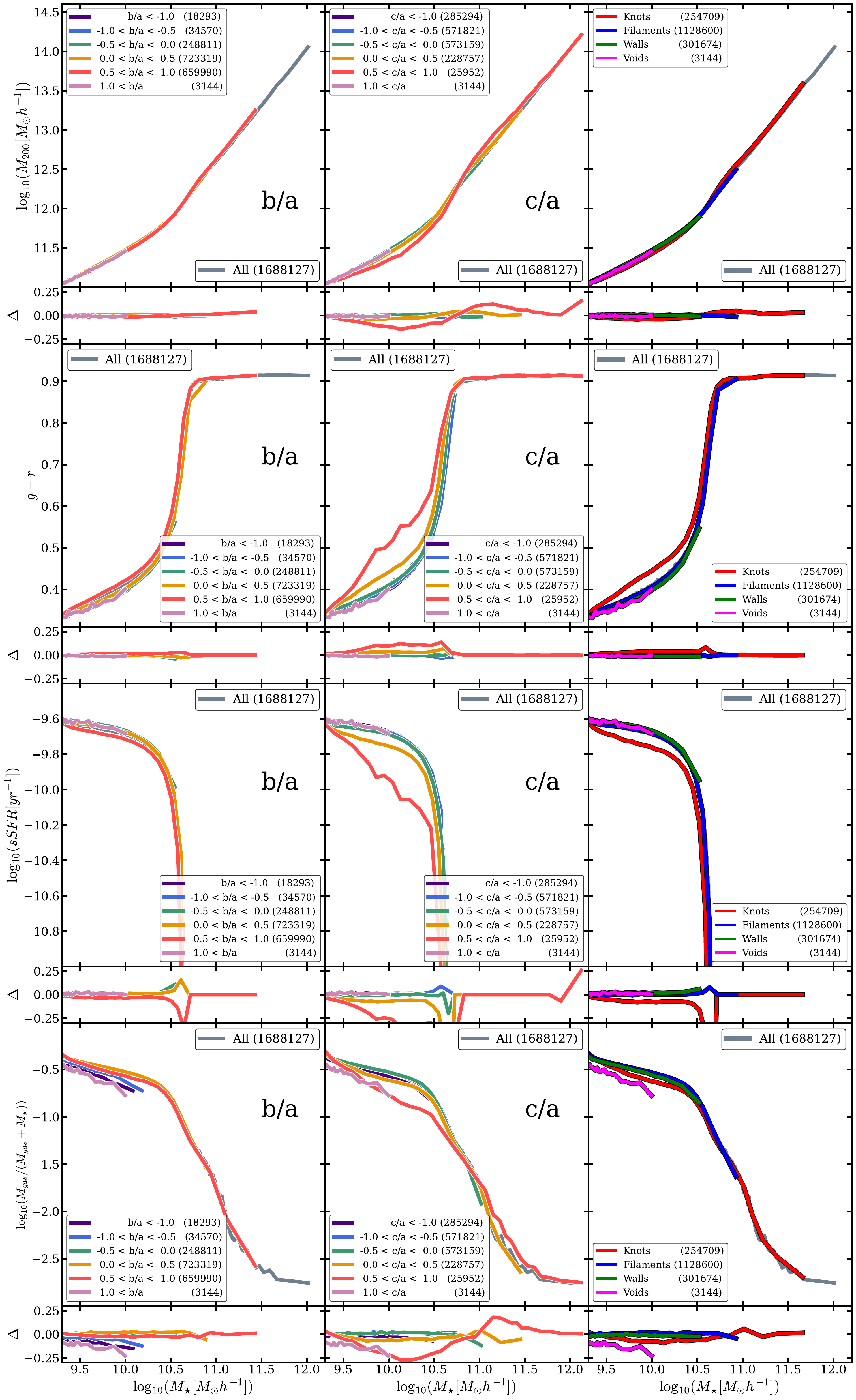}
    \caption{Galaxy properties as a function of the logarithm of the stellar mass. 
    From top to bottom: Logarithm of DM halo mass, color ({\em g-r}), 
    logarithm of the sSFR, and logarithm of the gas fraction.
    Each column displays the dependence of these properties on the environment: 
    equal bins in $b/a$ (left panels) and equal bins in $c/a$ (middle panels). 
    Each solid colored line corresponds to one of these bins, as indicated by the 
    legends in each panel. 
    Next to each bin, the number of galaxies contained within that 
    bin is specified. 
    The solid gray line in each panel represents 
    the total sample. 
    For comparison, the right panels show the dependence of these properties on 
    the environment according to the traditional method, categorizing 
    the environment into four types: knots, filaments, 
    walls, and voids.
    Each main panel is accompanied by a lower panel that shows the 
    difference ($\Delta$) between each colored line and the gray line.}
    \label{fig:prop_smass_ba_ca}
\end{figure*}

In this work we studied the influence of the environment 
on the main properties of galaxies in a continuous way, 
by analyzing the eigenvalues ratios estimated from the distribution 
of matter in the environment. 
Given the well-known dependence of galaxy properties on their 
stellar mass, in Fig.~\ref{fig:prop_smass_ba_ca} we analyze how 
the environment in which a galaxy resides affects this dependence. 
Each row shows a galaxy property as a function of 
the stellar mass \mbox{($M_{\star} [M_{\odot} h^{-1}]$)}, 
from top to bottom: DM halo mass 
\mbox{($M_{200} [M_{\odot} h^{-1}]$)}, color (g-r), 
sSFR (\emph{sSFR}$[yr^{-1}]$), and gas fraction
($M_{\rm gas}/(M_{\rm gas}+M_{*})$).
Left panels display the dependence of these properties with the 
environment expressed in terms of equal bins of $b/a$, 
while middle panels show them in equal bins of $c/a$. 
Each solid colored line corresponds to one of these bins, 
as indicated by the legends in each panel, 
with the number of galaxies in the bin in parentheses. 
As $b/a$ and $c/a$ increase, approaching 1, they indicate galaxies 
located in increasingly dense environments. 
However, the bins corresponding to $b/a>1$ and $c/a>1$, 
indicated with pink color, correspond to low-density regions 
(\mbox{$\delta < 0$)}. 
This is why these latter bins contain very few galaxies, which are
distributed over a very narrow range of stellar mass, and their
behavior deviates from the trend observed for the rest of the bins.
The number of galaxies located in this environment is 
only the 0.2 per cent of the sample.
In addition to the solid lines in different colors, each panel also 
includes a solid gray line representing median values of 
the entire sample analyzed. 
This gray line is the same across all three panels in each row. 
Each main panel is accompanied by a lower panel that shows
the difference ($\Delta$) between each colored line and the 
gray line representing the total sample.
It is noteworthy that the dependence of galaxy properties on their 
environment, as analyzed through these ratios, is biparametric, 
jointly dependent on both $b/a$ and $c/a$. 
However, when comparing the left and middle panels for each property, 
the most significant variations occur when considering changes in $c/a$.
This suggests that the relationship between eigenvalue $c/a$ is the 
most sensitive indicator of the dependence of galaxy properties 
on their environment.
For comparison, the right panels show the traditional classification 
of environments, categorizing galaxies into four types: 
knots, filaments, walls, and voids. 
This classification is based on the criteria outlined in 
Section~\ref{TT}, with a threshold value of $\lambda_{th} = 0$. 
Using the eigenvalues of the environment surrounding each galaxy 
in our sample, we assigned each galaxy to one of these four categories.
\\

The upper panels of Fig.~\ref{fig:prop_smass_ba_ca} clearly 
reveal the well-known correlation between a galaxy’s stellar mass 
\mbox{($M_{\star}$)} and the mass of its DM halo \mbox{($M_{200}$)}: 
more massive galaxies tend to reside in more massive halos. 
Assessing the influence of the environment on this relation (as indicated by 
the differently colored curves) we find a dependence on the surrounding cosmic 
environment, particularly evident in the trends associated with the eigenvalue ratio $c/a$.
In the left panel, where the colored curves represent different values of the 
intermediate-to-major eigenvalues ratio $b/a$, there are no significant differences 
among the curves, suggesting that this relation is largely insensitive to variations in $b/a$. 
In contrast, the upper-middle panel, where the colored curves correspond to different values 
of the minor-to-major eigenvalues ratio $c/a$, reveals a clear environmental dependence. 
Low-stellar-mass galaxies, with \mbox{$\log_{10}(M_{\star} [M_{\odot} h^{-1}]) \le 10.8$}, 
tend to inhabit less massive DM halos when located in regions where 
\mbox{$0.5 < c/a < 1$} (red line). 
Conversely, galaxies with \mbox{$\log_{10}(M_{\star} [M_{\odot} h^{-1}]) > 10.8$} are found 
in more massive halos when residing in regions with \mbox{$0.5 < c/a < 1$} (red line). 
Adopting a fixed smoothing scale can affect the 
interpretation of environmental effects on galaxy–halo connections, potentially 
introducing scale-dependent features that are not of physical origin.
With a fixed smoothing length, the region probed varies with the halo mass, 
which results in different environmental scales sampled for low and high 
mass halos.
To assess the robustness of our findings, we repeated the analysis 
presented in Fig.~\ref{fig:prop_smass_ba_ca} using larger smoothing scales:  
\mbox{$2$ and $4 \,{\rm Mpc}\,h^{-1}$}, shown in 
Fig.~\ref{fig:prop_sm_ba_ca_2} and Fig.~\ref{fig:prop_sm_ba_ca_4}, respectively, 
in the Appendix. 
As previously noted, a marked change in behavior emerges around 
\mbox{$\log_{10}(M_{\star} [M_{\odot} h^{-1}]) \sim 10.8$} when using the 
original smoothing scale of \mbox{$1 \,{\rm Mpc} \,h^{-1}$}. 
However, this transition shifts to a higher stellar mass of 
\mbox{$\log_{10}(M_{\star} [M_{\odot} h^{-1}]) \sim 11.2$} when adopting a 
smoothing scale of \mbox{$2 \,{\rm Mpc} \,h^{-1}$} and disappears entirely when 
the largest scale is applied. 
At \mbox{$4 \,{\rm Mpc}\, h^{-1}$}, galaxies across the entire stellar mass range 
consistently exhibit the same trend, i.e., those residing in regions where 
\mbox{$0.5 < c/a < 1$} inhabit less massive halos than galaxies of the same 
stellar mass located in regions with smaller $c/a$ values.
When comparing this to the right panel, which uses traditional 
environmental classification, the same trend is observed but is notably 
more attenuated. 
Examining the differences shown in the bottom panels ($\Delta$) reveals that the 
variations are significantly more pronounced when considering $c/a$ 
rather than the traditional classification. 
This comparison underscores the added value of describing the environment 
through eigenvalue ratios, providing a complementary perspective to 
traditional methods: it offers a straightforward and continuous characterization 
that naturally reveals the dependence of galaxy properties on their 
surrounding environment.
It is also worth noting that the lack of a significant trend 
with $b/a$ is not unexpected. 
In regions where matter collapses nearly isotropically, 
$b/a$ is inherently constrained between $c/a$ and 1 
(i.e., \mbox{$c/a \le b/a \le 1$}), limiting its dynamical range. 
While $b/a$ might still hold some discriminatory power in differentiating 
between regions where matter collapses in one or two directions, 
our analysis reveals no strong signal across that interface, 
suggesting its role in tracing environmental effects is subdominant 
compared to $c/a$.
\\

When considering color and SFR, blue galaxies are typically 
lower-mass systems that are actively forming stars, whereas red galaxies 
tend to be more massive and have largely ceased star formation. 
The transition from blue to red galaxies occurs through a process known 
as quenching, where star formation declines resulting in a redder color.
The second row panels show the dependence of color ({\em g-r)}, 
where it is evident that more massive galaxies are redder, while less 
massive ones are bluer. 
Taking the environment into account, particularly focusing on $c/a$, 
there is a clear trend for galaxies of comparable stellar mass: 
for a given stellar mass, galaxies tend to become 
progressively redder as the values of $c/a$ increase toward unity, 
which corresponds to regions with $\delta > 0$.
This trend is noticeable for galaxies with 
\mbox{$\log_{10}(M_{\star} [M_{\odot} h^{-1}]) < 10.8$}, 
but becomes negligible for more massive galaxies. 
Concerning the sSFR shown in the 
third-row panels, it displays a pattern similar to that of color 
and admits a similar analysis: 
galaxies located in regions where the $c/a$ value is close to 1 
(with \mbox{$\delta > 0$}), tend to have lower SFRs, 
suggesting they are more quiescent 
compared to galaxies of similar stellar mass found in regions with lower 
$c/a$ values.
This trend is again observed for galaxies with 
\mbox{$\log_{10}(M_{\star} [M_{\odot} h^{-1}]) < 10.8$}.
Finally, we analyzed the gas fraction of galaxies 
and its dependence on the environment 
(bottom panels). 
The relationship between gas fraction and stellar mass is closely linked to 
SFR and galaxy color. 
It is well established that lower-mass galaxies have a higher gas fraction available 
for star formation, which also contributes to their bluer color. 
In contrast, more massive galaxies have less gas available for star formation, 
making them redder and more quiescent.
When examining the environmental dependence, a trend similar to that observed 
for galaxy color and sSFR emerges: 
lower-mass galaxies exhibit a reduced gas fraction in regions 
where the $c/a$ values increase and approach unity, which corresponds to regions 
with $\delta > 0$.
However, unlike the trend seen for color and sSFR, where more massive galaxies 
show no significant environmental variation, in the case of gas fraction, 
more massive galaxies reverse this pattern, displaying a higher gas fraction 
when $c/a$ value becomes closer to 1 (\mbox{$\delta > 0$}). 
As previously discussed when analyzing the variation in the dependence 
of halo mass on stellar mass according to the environment in which galaxies 
reside, the observed change in the trend at the high-mass end is not a 
physical effect but rather a consequence of using a smoothing scale 
that is small relative to the size of the most massive galaxies. 
When the smoothing scale is increased, this trend disappears. 
Using a smoothing scale of \mbox{$4 \,{\rm Mpc}\,h^{-1}$}, galaxies located 
in regions where $c/a$ approaches 1 (\mbox{$\delta > 0$}) 
exhibit lower gas fractions than galaxies with 
the same stellar mass located in regions with smaller $c/a$ values. 
This trend is consistent across the entire range of stellar masses analyzed 
(please, see Fig.~\ref{fig:prop_sm_ba_ca_4} in the Appendix).
This trend is also evident in the right panel, although it is much less pronounced.
\\

In summary, the continuous description of the cosmic web offers a 
complementary and detailed perspective on the environmental dependence of galaxy properties. 
Its continuous nature allows for the identification of subtle trends and smooth variations.
Notably, the mean trends observed using $c/a$ align with those obtained through 
the traditional classification, indicating that our method is consistent 
with previous findings while offering enhanced detail.
\\

\begin{figure*}[h!]
    \centering
    \includegraphics[width=0.73\textwidth]{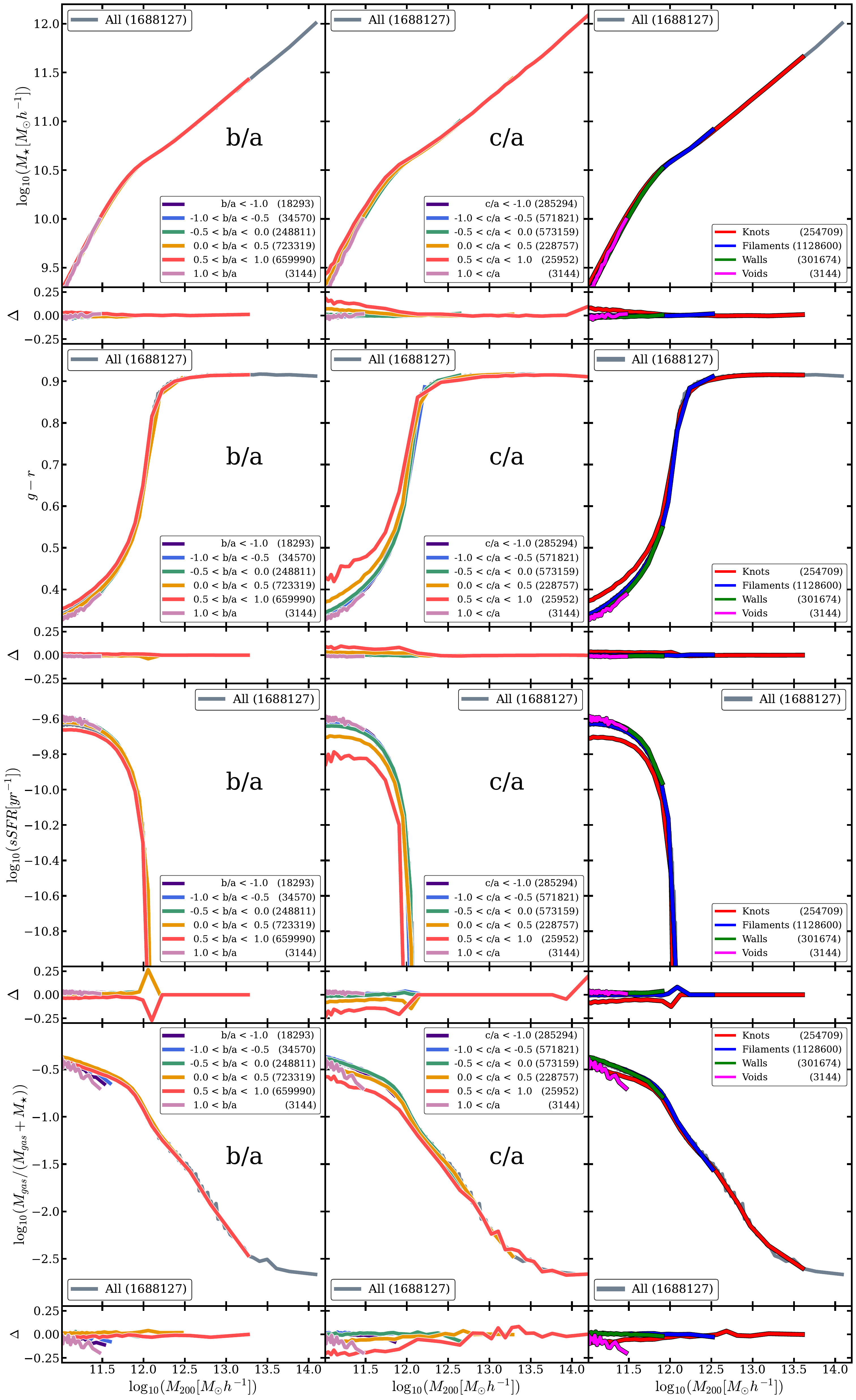}
    \caption{Similar to Fig.~\ref{fig:prop_smass_ba_ca} but as a function 
    of the logarithm of the DM halo mass instead of the stellar 
    mass of the galaxies.}
    \label{fig:prop_M200_ba_ca}
\end{figure*}

For completeness, we present Fig.~\ref{fig:prop_M200_ba_ca}, which is similar 
to Fig.~\ref{fig:prop_smass_ba_ca} but shows galaxy properties as a function 
of DM halo mass (\mbox{$M_{200}$}) instead of stellar mass. 
Comparing these figures, it is evident that they display very similar trends. 
As in Fig.~\ref{fig:prop_smass_ba_ca}, all properties show a more pronounced 
dependence on the environment when considering the eigenvalue ratio $c/a$.
Comparing galaxies residing in DM halos of similar mass, those 
located in environments with $c/a$ close to 1 
(\mbox{$\delta > 0$}) are redder, exhibit 
lower SFRs, and have a lower gas fraction. 
This environmental dependence is significant only for galaxies whose DM 
halos are less massive than \mbox{$ M_{200} = 10^{12}$}.
\\

\section{Conclusions} \label{Conclusions}

In this work we present a new continuous description of the cosmic web 
using the ratios of the eigenvalues of the Hessian matrix derived from the 
matter distribution. 
We applied this method to a numerical simulation coupled with a semi-analytic 
model of galaxy formation and analyzed the dependence of galaxy properties 
on the environment.
\\

Galaxy properties show a clear dependence on the surrounding environment. 
More massive galaxies, which tend to be redder, quenched, and gas-poor, 
are most commonly found in regions where the eigenvalues $a$, $b$, and $c$ 
are roughly equal. 
In contrast, lower-mass galaxies, typically bluer, star-forming, and gas-rich, 
are more likely to reside in environments 
where the values of $b/a$ and $c/a$ decrease. 
While the properties of the most massive galaxies show little sensitivity to 
environmental factors, low-mass galaxies are significantly more influenced 
by their surroundings. 
Although the environmental dependence of galaxy properties is biparametric, 
as it is jointly dependent on both $b/a$ and $c/a$, it is predominantly 
driven by the eigenvalue ratio $c/a$.
\\

We compared our results with those obtained using the traditional method 
of classifying environments into four categories: knots, filaments, walls, and voids. 
Our new eigenvalue ratio method offers two key advantages.
First, it eliminates the need for arbitrary thresholds to define each environmental 
category. 
Second, the dependence of galaxy properties on the environment is much more 
pronounced when directly considering the eigenvalue ratio, compared to the four 
traditional categories.
This is because the dependence becomes particularly strong 
in environments where the values of $c/a$ approach unity 
in regions with \mbox{$\delta > 0$}. 
In this regime, the three principal axes of the matter distribution are nearly equal, 
indicating a more symmetric and isotropic collapse. 
However, when all such environments are indiscriminately grouped into a single category 
(knots), the subtle differences associated with varying degrees of collapse symmetry 
are effectively averaged out, which dilutes the underlying trend.
\\

Furthermore, using eigenvalue ratios enables the study of the temporal evolution 
of galaxy environments without introducing arbitrary thresholds that may vary 
over time, allowing for more direct and accurate comparisons across different epochs. 
These findings extend beyond the scope of this paper and will be presented in a 
forthcoming study.
\\

\begin{acknowledgements}
We would like to thank the referee for carefully reading the manuscript and 
providing many comments and suggestions that significantly improved our paper.
The SMDPL simulation was performed at LRZ Munich within the pr87yi project. 
The authors gratefully acknowledge funding for this project from the 
Gauss Centre for Supercomputing e.V.
(www.gauss-centre.eu) by providing computing time 
on the GCS Supercomputer SUPERMUC-NG at the Leibniz Supercomputing Centre 
(www.lrz.de).
The CosmoSim database (www.cosmosim.org) used in this paper is a service of the 
Leibniz Institute for Astrophysics Potsdam (AIP). 
This work has been partially supported by the Consejo de
Investigaciones Cient\'ificas y T\'ecnicas de la Rep\'ublica Argentina
\mbox{(CONICET)}, Secretar\'ia de Ciencia y T\'ecnica de la Universidad Nacional 
de C\'ordoba (SeCyT) and Agencia Nacional de Promoci\'on Cient\'ifica y 
Tecnol\'ogica (PICT 2019-1600), Argentina. 
GY  would like to thank Ministerio de Ciencia e Innovaci\'on (Spain) for 
financial support under the project grant PID2021-122603NB-C21. 
MGA and GY also thanks  European Union Horizon 2020 Research and Innovation 
Programme  for partial financial support under the Marie Sklodowska-Curie 
grant agreement No 734374-LACEGAL. 
MGA acknowledges the hospitality of the Departamento de F\'isica Te\'orica at 
the Universidad Autónoma de Madrid during a scientific visit 
when part of this project was carried out.

\end{acknowledgements}

\bibliographystyle{aa}
\bibliography{Bibliography.bib} 

\begin{appendix}

\section{Analyzing different smoothing scales} \label{App}

\begin{figure*}
    \centering
    \begin{subfigure}[b]{0.5\textwidth}
        \centering
        \includegraphics[width=\textwidth]{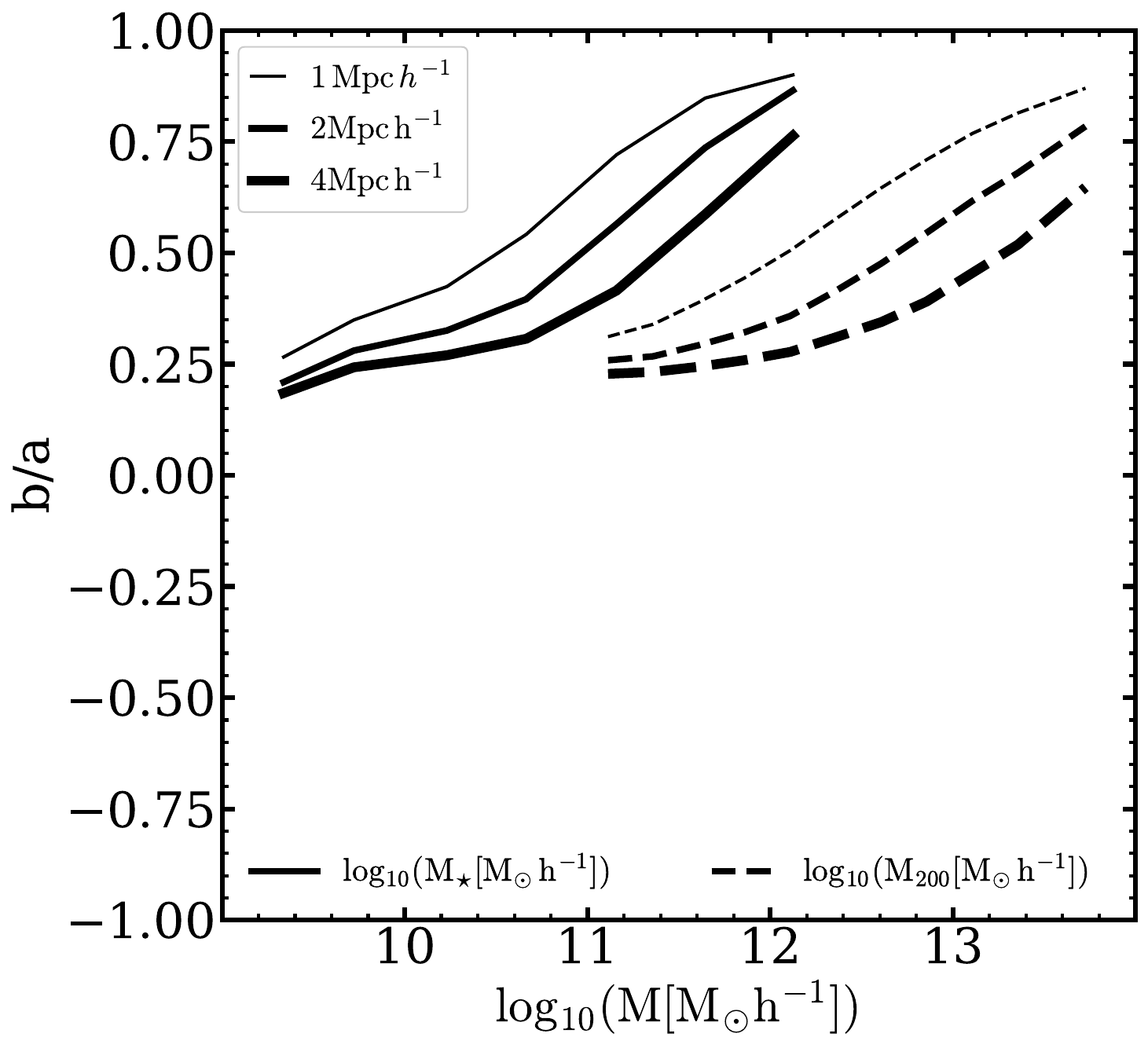}
    \end{subfigure}%
    \begin{subfigure}[b]{0.5\textwidth}
        \centering
        \includegraphics[width=\textwidth]{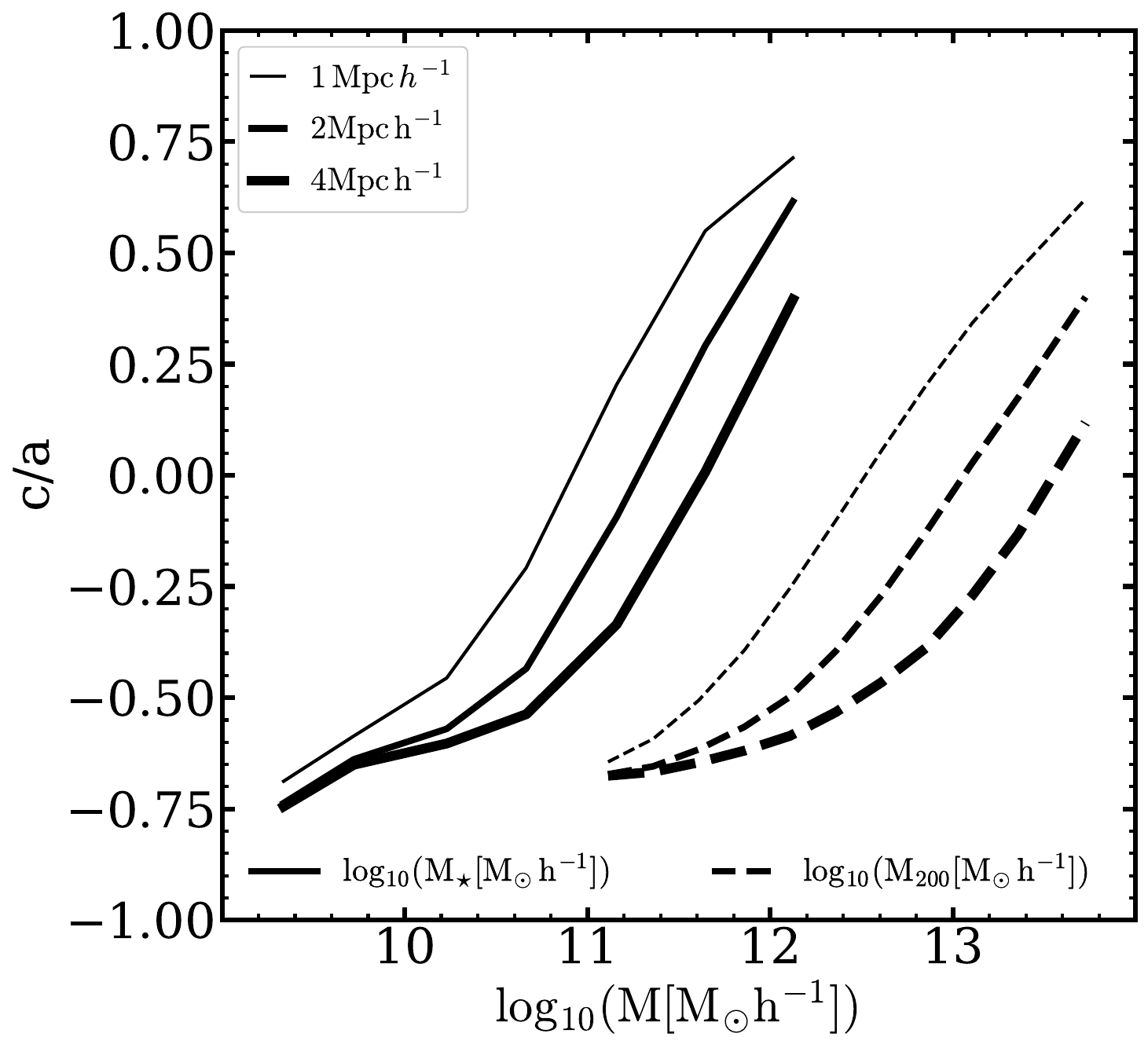}
    \end{subfigure}
    \vspace{0.3em}
    \caption{Eigenvalues ratios $b/a$ (left panel) and $c/a$ (right panel) 
    as a function of the logarithm of stellar mass (solid black lines) and the 
    logarithm of DM halo mass (dashed black lines) for the three 
    smoothing scales analyzed \mbox{(1, 2, and 4 $\,{\rm Mpc}\,h^{-1}$)} 
    differentiated by the thickness of the lines, as indicated in the legend.
    Eigenvalues are estimated from the tidal tensor.}
    \label{fig:T_smoothings}
\end{figure*}

\begin{figure*}
    \centering
    \begin{subfigure}[b]{0.5\textwidth}
        \centering
        \includegraphics[width=\textwidth]{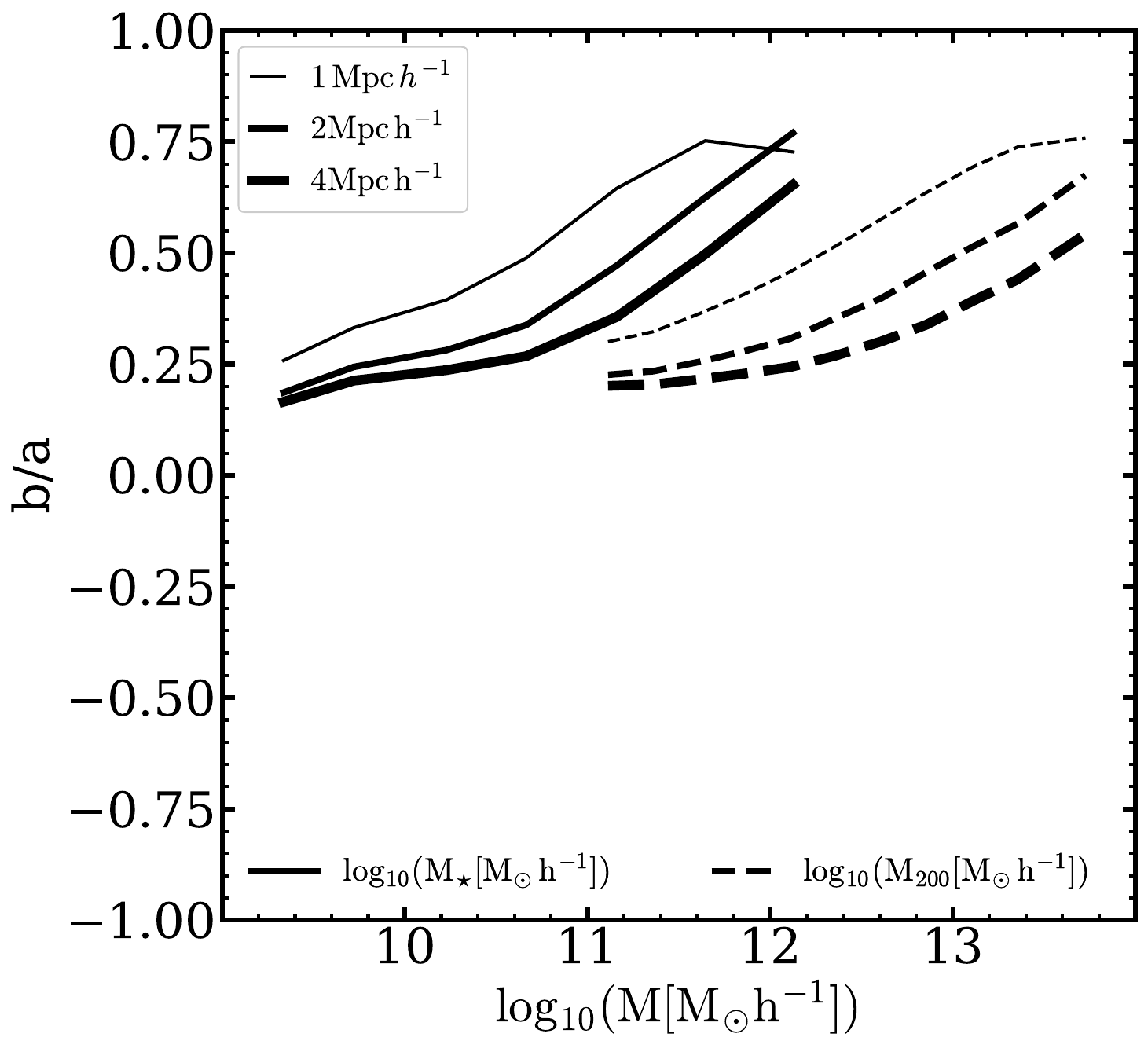}
    \end{subfigure}%
    \begin{subfigure}[b]{0.5\textwidth}
        \centering
        \includegraphics[width=\textwidth]{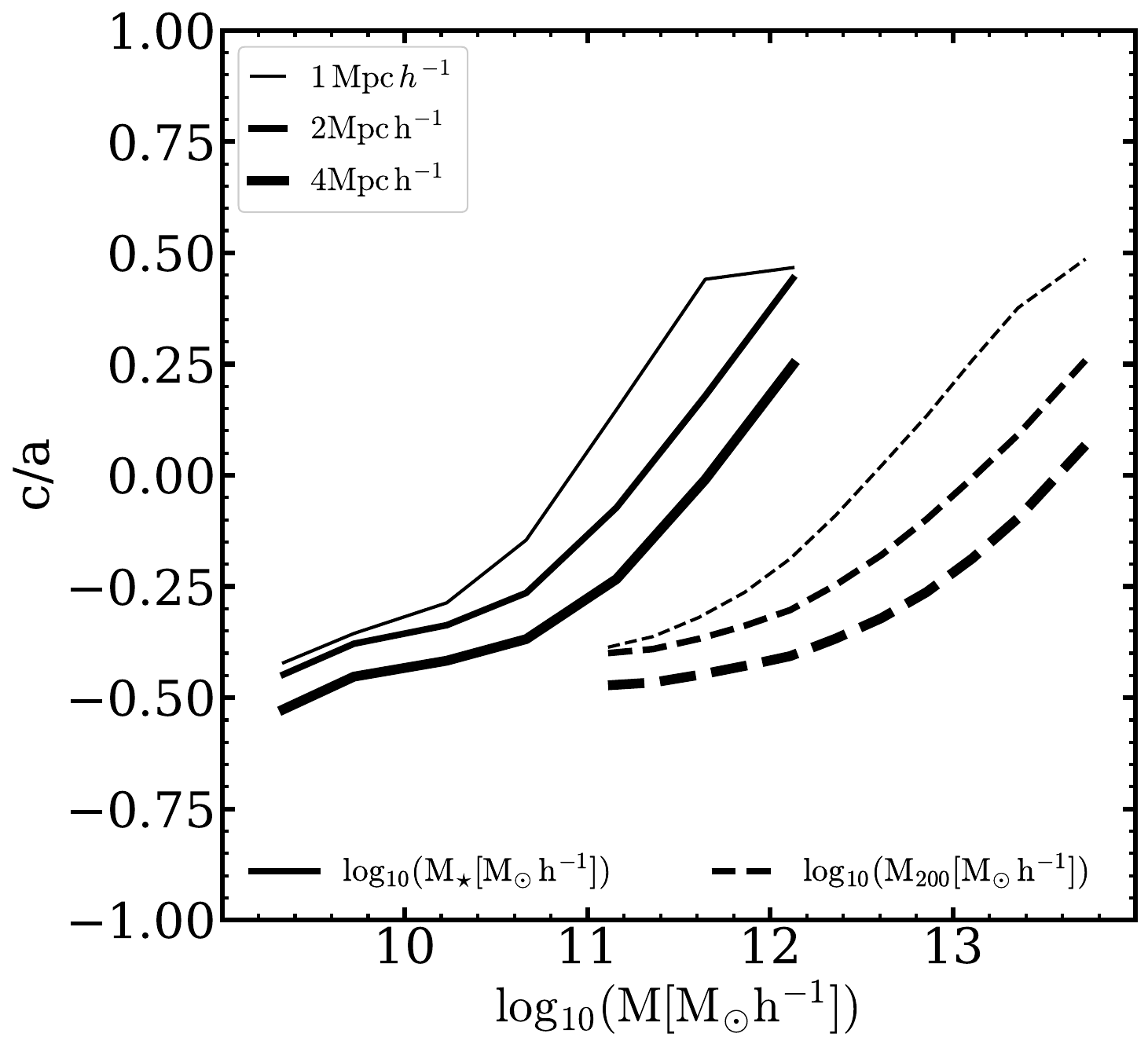}
    \end{subfigure}
    \vspace{0.3em} 
    \caption{Eigenvalues ratios $b/a$ (left panel) and $c/a$ (right panel) 
    as a function of the logarithm of stellar mass (solid black lines) and the 
    logarithm of DM halo mass (dashed black lines) for the three 
    smoothing scales analyzed \mbox{(1, 2, and 4 $\,{\rm Mpc}\,h^{-1}$)} 
    differentiated by the thickness of the lines, as indicated in the legend.
    Eigenvalues are estimated from the shear velocity field.}
    \label{fig:V_smoothings}
\end{figure*}

The influence of the smoothing scale used to categorize the cosmic web 
has been extensively analyzed in numerous studies
(e.g., \citealt{Aycoberry:2024, Zhang:2024, Balaguera:2024, 
Pfeifer:2022, Wang:2020}, among others). 
These works demonstrate that the outcomes of different analyses are 
significantly affected not only by the chosen smoothing scale but 
also by the choice of threshold used to classify the cosmic web, 
depending on how many eigenvalues exceed that value. 
The extent of these dependences varies depending 
on the specific properties or phenomena under investigation.
In this appendix, we demonstrate that our results remain robust under 
variations in the smoothing scale \mbox{(1, 2, and 4 $\,{\rm Mpc}\,h^{-1}$)} 
and when the shear velocity field is used instead of the tidal field 
to estimate the eigenvalues a, b and c.
\\

Figure~\ref{fig:T_smoothings} illustrates the variation in the median values of 
the ratio of eigenvalues $b/a$ (left panel) and $c/a$ (right panel) as a 
function of the stellar mass (solid black lines) and the halo mass 
(dashed black lines) for the three smoothing scales 
analyzed \mbox{(1, 2, and 4 $\,{\rm Mpc}\,h^{-1}$)} differentiated by the 
thickness of the lines, as indicated in the legend.
In these figures, as in all figures presented throughout the paper, the eigenvalues 
a, b and c were computed from the tidal field, derived from the 
density field of DM particles. 
For comparison, Fig.~\ref{fig:V_smoothings} is equivalent to 
Fig.~\ref{fig:T_smoothings}, but in this case, the eigenvalues a, b and c 
were estimated from the shear velocity field.
\\

From these figures, we conclude that although increasing the 
smoothing scale does have an impact, particularly in the higher mass bins, 
the overall trends persist, and our main conclusions remain robust.
In our proposed method for describing the cosmic web, the c/a ratio emerges 
as the most effective parameter for capturing the relationship between 
galaxy properties and their surrounding environment. 
This dependence is remarkably clear and continuous across 
the parameter space, enabling a detailed and nuanced exploration of how galaxy 
properties vary with their environment.
\\

\begin{figure*}[h!]
    \centering
    \includegraphics[width=0.73\textwidth]{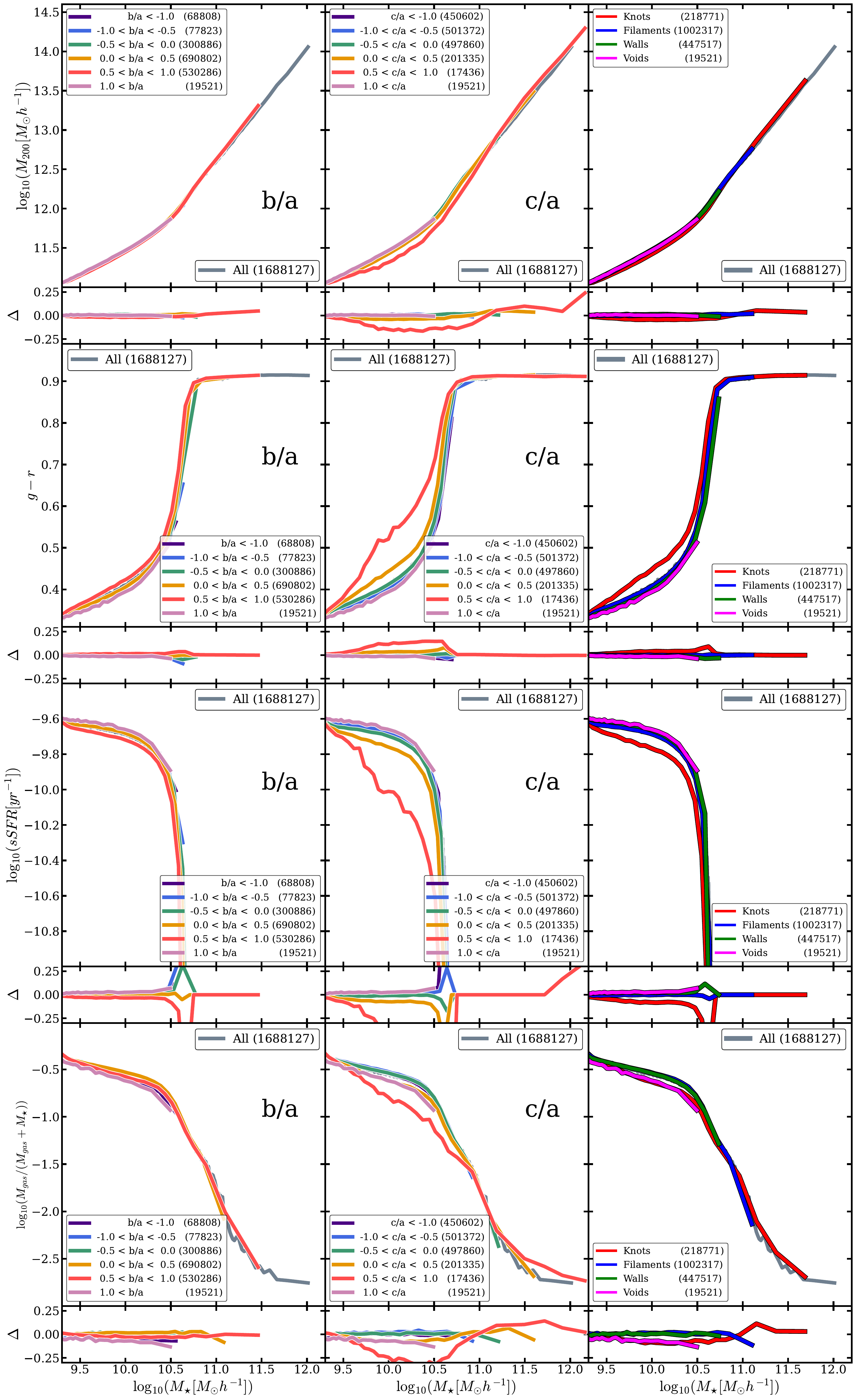}
    \caption{Similar to Fig.~\ref{fig:prop_smass_ba_ca} but using a smoothing scale 
    equals to 2 $\,{\rm Mpc}\,h^{-1}$.}
    \label{fig:prop_sm_ba_ca_2}
\end{figure*}

\begin{figure*}[h!]
    \centering
    \includegraphics[width=0.73\textwidth]{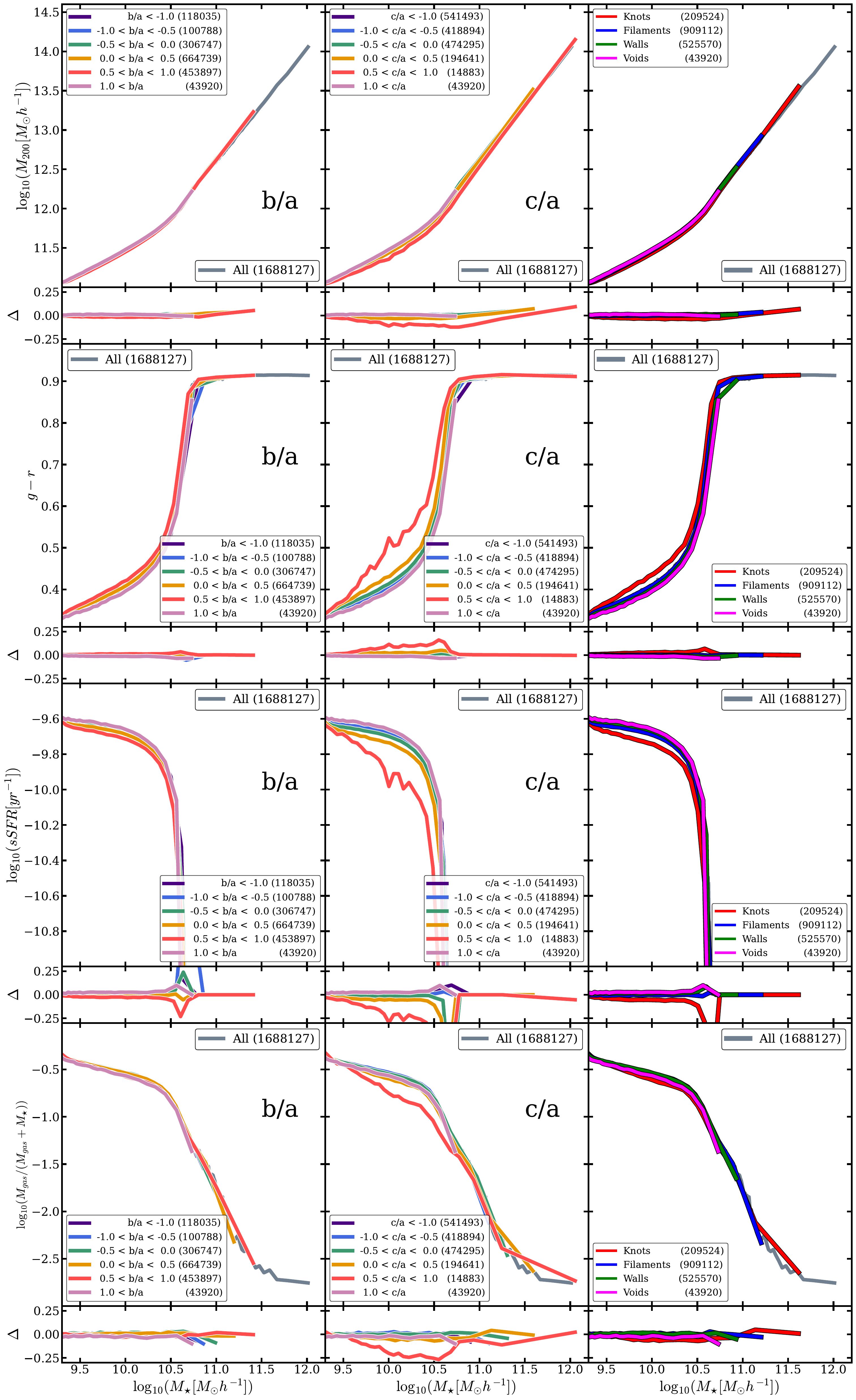}
    \caption{Similar to Fig.~\ref{fig:prop_smass_ba_ca} but using a smoothing scale 
    equals to 4 $\,{\rm Mpc}\,h^{-1}$.}
    \label{fig:prop_sm_ba_ca_4}
\end{figure*}

The choice of smoothing scale used to define the cosmic web constitutes 
a well-known limitation affecting many of the currently employed methods, including 
the traditional approach that classifies the cosmic web into four distinct 
components \citep[][among others] {Hahn:2007a, Sousbie:2008, Forero-Romero:2009, 
Libeskind:2013a, Libeskind2018, Wang:2020, Jaber:2024, Zhang:2024}. 
Recognizing the relevance of the smoothing scale in characterizing the environment 
in which galaxies reside—and the potential impact this choice may have on the results 
presented in this work—we repeated our analysis of the dependence of galaxy properties 
on their surrounding environment, adopting different smoothing lengths for the 
environmental description.
Figs.~\ref{fig:prop_sm_ba_ca_2} and~\ref{fig:prop_sm_ba_ca_4} 
present galaxy properties as a function of stellar mass, in the same 
manner as Fig.~\ref{fig:prop_smass_ba_ca}. 
While Fig.~\ref{fig:prop_smass_ba_ca} corresponds to a smoothing length of 
\mbox{1 $\,{\rm Mpc}\,h^{-1}$}, Figs.~\ref{fig:prop_sm_ba_ca_2} and~\ref{fig:prop_sm_ba_ca_4} 
display the results obtained with smoothing 
lengths of \mbox{2 and 4 $\,{\rm Mpc}\,h^{-1}$}, respectively.
A comparative analysis of the corresponding panels across the three figures—each 
based on a different smoothing scale—reveals that the most significant trends are 
preserved regardless of the chosen scale.
The most notable difference arises in the dependence of galaxy properties on 
the $c/a$ parameter. 
As discussed in Section~\ref{GP}, galaxies exhibit a stronger correlation 
with $c/a$ than with $b/a$. 
Focusing on the variation of halo mass ($M_{200}$) with $c/a$ (middle upper panels), 
we find that when the environment is defined using a smoothing scale of 
\mbox{1 $\,{\rm Mpc}\,h^{-1}$}, low-stellar-mass galaxies—up to 
\mbox{$\log_{10}(M_{\star} [M_{\odot} h^{-1}]) \sim 10.8$}—tend to reside in less 
massive DM halos when the values of $c/a$ rise and 
tend toward unity (red line). 
Conversely, for galaxies with \mbox{$\log_{10}(M_{\star} [M_{\odot} h^{-1}]) > 10.8$}, 
the trend reverses: they are found in more massive halos when located in environments 
where the values of $c/a$ increase, approaching unity (red line). 
As the smoothing scale increases, the stellar mass at which this reversal occurs also 
shifts to higher values. 
Specifically, this transition occurs at \mbox{$\log_{10}(M_{\star} [M_{\odot} h^{-1}]) \sim 11.2$} 
when using a smoothing scale of \mbox{2 $\,{\rm Mpc}\,h^{-1}$}, and at even higher 
stellar masses for the \mbox{4 $\,{\rm Mpc}\,h^{-1}$} scale.
For the other galaxy properties analyzed, no significant differences are observed 
with varying smoothing scales. 
\\

\begin{figure*}[h!]
    \centering
    \includegraphics[width=0.73\textwidth]{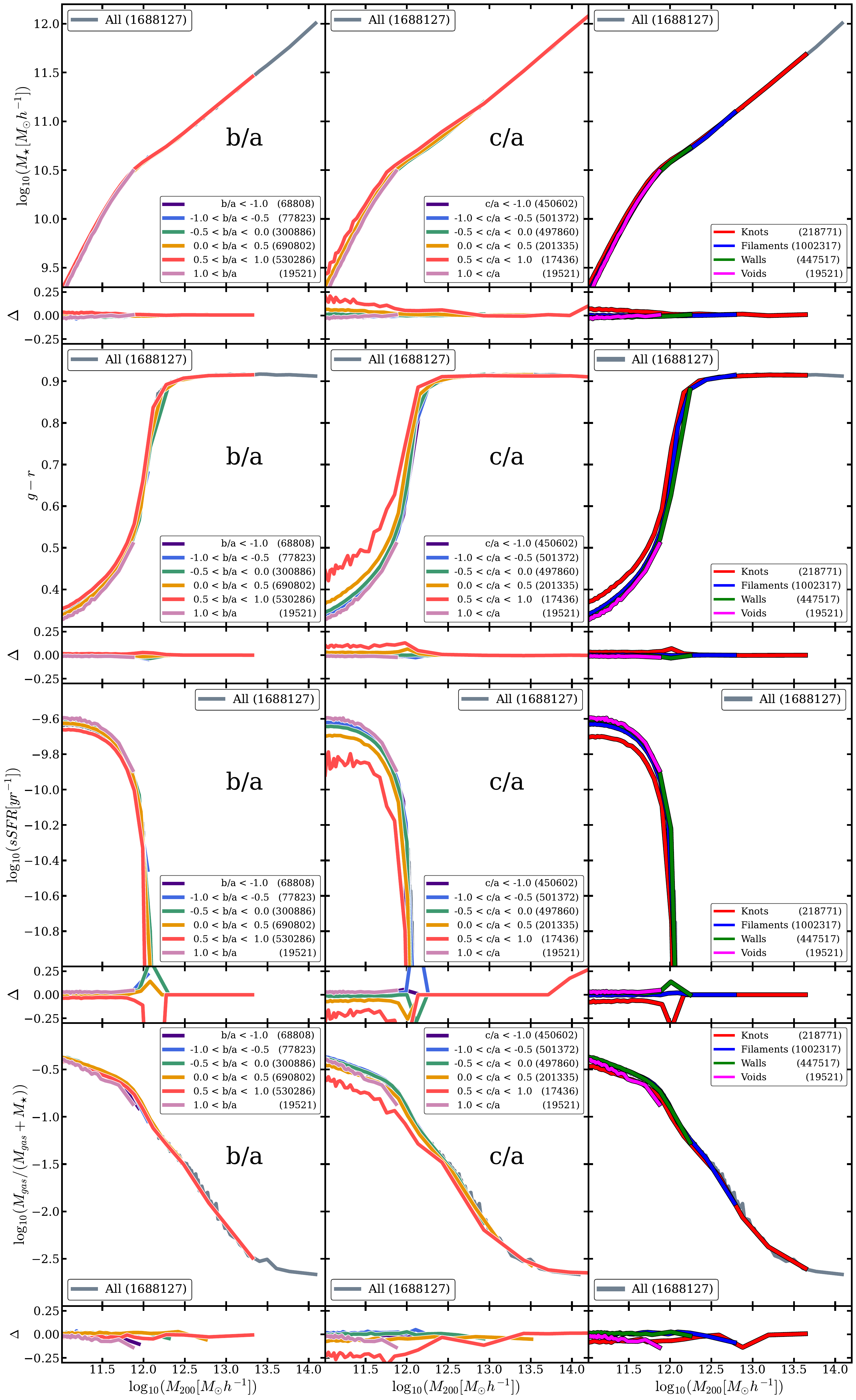}
    \caption{Similar to Fig.~\ref{fig:prop_M200_ba_ca} but using a smoothing scale 
    equals to 2 $\,{\rm Mpc}\,h^{-1}$.}
    \label{fig:prop_M200_ba_ca_2}
\end{figure*}

\begin{figure*}[h!]
    \centering
    \includegraphics[width=0.73\textwidth]{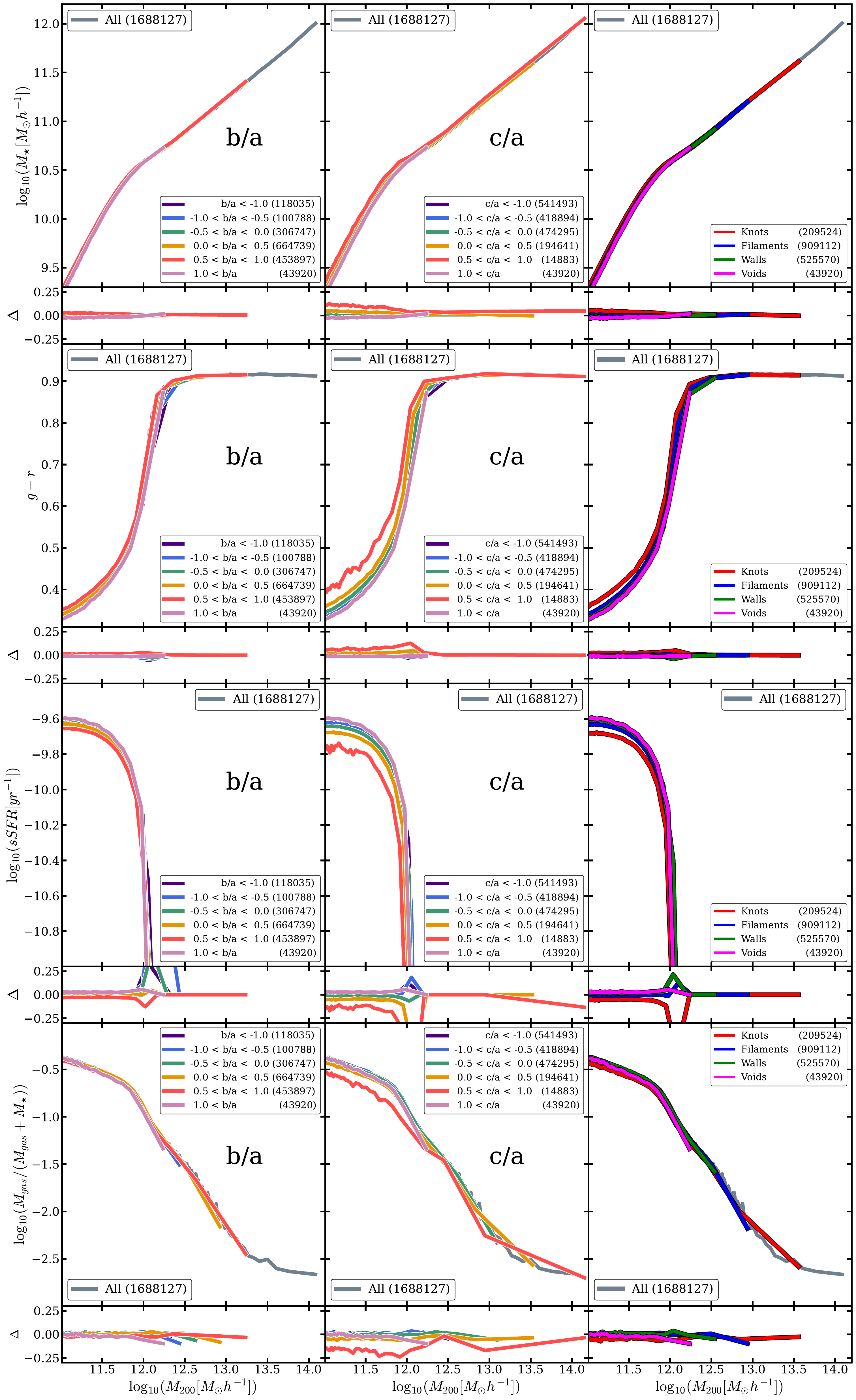}
    \caption{Similar to Fig.~\ref{fig:prop_M200_ba_ca} but using a smoothing scale 
    equals to 4 $\,{\rm Mpc}\,h^{-1}$.}
    \label{fig:prop_M200_ba_ca_4}
\end{figure*}

For completeness, we also include Figs~\ref{fig:prop_M200_ba_ca_2}
and~\ref{fig:prop_M200_ba_ca_4}, which are analogous to Fig.~\ref{fig:prop_M200_ba_ca} 
(where astrophysical properties are analyzed as a function of the DM halo mass) 
but instead define the galaxy environment using smoothing scales of 
\mbox{2 and 4 $\,{\rm Mpc}\,h^{-1}$}, respectively.
In this case as well, we find no significant differences in the results when changing 
the smoothing length used to define the environment of galaxies.
In conclusion, as anticipated in Section~\ref{TT}, the main results presented in this 
study remain robust across different choices of smoothing length in the definition 
of the environment.
\\

\end{appendix}

\end{document}